\documentclass[twocolumn,superscriptaddress,showpacs,prd,aps,amsmath,amssymb,nofootinbib]{revtex4-1}

\usepackage{graphicx}
\usepackage{longtable}

\newcommand{\beq}{\begin{equation}} 
\newcommand{\eeq}{\end{equation}} 
\newcommand{\beqn}{\begin{eqnarray}} 
\newcommand{\eeqn}{\end{eqnarray}}

\newcommand{\zD}{{\raise1.0ex\hbox{${}^{\ \circ}$}}\!\!\!\!\!D}
\newcommand{\alone}{{\raise0.5ex\hbox{${}^{\ 1}$}}\!\!\!\!\alpha}

\newcommand{\nalam}{\mathrel{\raise0.9ex\hbox{$^\lambda$}\mkern-14mu
\lower0.0ex\hbox{$\nabla$}}}

\newcommand{\zeroD}{{\raise1.0ex\hbox{${}^{\ \circ}$}}\!\!\!\!\!D}
\newcommand{\zLap}{{\raise1.0ex\hbox{${}^{\ \circ}$}}\!\!\!\!\Delta}
\newcommand{\zna}{{\raise1.0ex\hbox{${}^{\ \circ}$}}\!\!\!\!\!\nabla}
\newcommand{\zS}{{\raise1.0ex\hbox{${}^{\ \circ}$}}\!\!\!\!\!S}

\usepackage[normalem]{ulem}
\usepackage{epstopdf}
\usepackage{color}
\usepackage{soul}
\usepackage{ulem}
\usepackage{bm}
\usepackage{xspace}

\usepackage[]{amsmath}
\usepackage[]{amsfonts}
\usepackage[]{amssymb}
\usepackage[]{mathrsfs}
\usepackage[]{mathtools}
\usepackage{times}
\usepackage{hyperref}
\hypersetup{
  colorlinks=true,        
  linkcolor=blue,         
  citecolor=cyan,         
}
\newcommand{\apjl}{Astrophys. J. Lett.}

\newcommand{\aap}{Astron. Astrophys.}
\newcommand{\mnras}{Mon. Not. R. Astron. Soc.}

\def\QEQ{{%
			\setbox0\hbox{$I$}%
			\rlap{\hbox to \wd0{\hss--\hss}}\box0
		}}

\begin{document}

\title{The precursor of GRB211211A: a tide-induced giant quake?}

\author{Enping Zhou}
\affiliation{Huazhong University of Science and Technology, School of Physics, 1037 Luoyu Road, Wuhan, 430074, China}

\author{Yong Gao}
\affiliation{Kavli Institute for Astronomy and Astrophysics, Peking University, Beijing 100871, China}
\affiliation{School of Physics, Peking University, Beijing 100871, China}

\author{Yurui Zhou}
\affiliation{Huazhong University of Science and Technology, School of Physics, 1037 Luoyu Road, Wuhan, 430074, China}

\author{Xiaoyu Lai}
\affiliation{Department of Physics and Astronomy, Hubei University of Education, Wuhan 430205, China}
\affiliation{Research Center for Astronomy, Hubei University of Education, Wuhan 430205, China}

\author{Lijing Shao}
\affiliation{Kavli Institute for Astronomy and Astrophysics, Peking University, Beijing 100871, China}

\author{Weiyang Wang}
\affiliation{University of Chinese Academy of Sciences, Beijing 100049, China}
\affiliation{School of Physics, Peking University, Beijing 100871, China}

\author{Shaolin Xiong}
\affiliation{Key Laboratory of Particle Astrophysics, Institute of High Energy Physics, Chinese Academy of
Sciences, 19B Yuquan Road, Beijing 100049, China}

\author{Renxin Xu}
\affiliation{School of Physics, Peking University, Beijing 100871, China} \affiliation{Kavli Institute for Astronomy and Astrophysics, Peking University, Beijing 100871, China}

\author{Shuxu Yi}
\affiliation{Key Laboratory of Particle Astrophysics, Institute of High Energy Physics, Chinese Academy of
Sciences, 19B Yuquan Road, Beijing 100049, China}

\author{Han Yue}
\affiliation{School of Earth and Space Sciences, Peking University, Beijing 100871, China}

\author{Zhen Zhang}
\affiliation{Key Laboratory of Particle Astrophysics, Institute of High Energy Physics, Chinese Academy of
Sciences, 19B Yuquan Road, Beijing 100049, China}

\date{\today}

\begin{abstract}
The equilibrium configuration of a solid strange star in the final inspiral phase with another compact object is generally discussed, and the starquake-related issue is revisited, for a special purpose to understand the precursor emission of binary compact star merger events (e.g., that of GRB211211A).
As the binary system inspirals inward due to gravitational wave radiation, the ellipticity of the solid strangeon star increases due to the growing tidal field of its compact companion. 
Elastic energy is hence accumulated during the inspiral stage which might trigger a starquake before the merger when exceeds a critical value.
The energy released during such starquakes is calculated and compared to the precursor observation of GRB211211A.
The result shows that the energy might be insufficient for binary strangeon-star case unless the entire solid strangeon star shatters, and hence favors a black hole-strangeon star scenario for GRB211211A.
The timescale of the precursor as well as the frequency of the observed quasi-periodic-oscillation have also been discussed in the starquake model. 

\end{abstract}

\maketitle

\section{Introduction}

It is well known that the puzzling nature of pulsar's interior is essentially relevant to the fundamental strong interaction at low-energy scale, the challenging non-perturbative quantum chromo-dynamics (or strong QCD~\citep{2009RvMP...81.1773E}), but this unknown state could be the first big problem to be solved in the era of multi-messenger astronomy~\citep{2019PrPNP.10903714B}.
Besides the conventional neutron star (NS) model, pulsars are proposed alternatively to be solid strange stars~\citep{Xu2003} (or strangeon stars), and we are then developing a strangeon star model in order to understand extreme and mysterious events in astrophysics.
Certainly, quakes can naturally occur on solid strangeon stars, and a giant quake model has already been proposed for the super-flares of isolated soft $\gamma$-ray repeaters~\citep{2006MNRAS.373L..85X}.
Futhermore, tide-induced quakes in binary have also been discussed, appearing as so-called a sudden change in
the tidal deformability at a certain breaking frequency of gravitational wave (GW)~\citep{2019EPJA...55...60L}.
Can tide-induced quakes be manifest in the electromagnetic (EM) wave? This is our focus here, and luckily, the precursor of GRB211211A could be a typical example.
In the future, more such events, especially combined with the LVK-O4 observing run (e.g., \citep{2023arXiv230204147C}), would surely be expected.

Without doubt, the observation of GW170817~\citep{Abbott2017} together with its EM counterparts GRB170817A and AT2017gfo~\citep{Abbott2017b} has announced the birth of the multi-messenger astronomy era. This event has largely enriched our knowledge on the nature of short gamma-ray bursts (sGRB) ~\citep{Abbott2017d,Narayan92}, the state of matter at supranuclear densities~\citep{Ruiz2017,Rezzolla2017,Bauswein2017b,Margalit2017,Shibata2019,Kiuchi2019} as well as the origin of heavy elements in the Universe~\citep{Abbott2017c,Eichler89}. The EM counterparts have been detected in almost every band from radio to gamma ray, however, these observations all happen during the post-merger phase. As an implementation, EM signals prior to the merger (i.e., the precursor observation), if detected, could significantly improve our understanding of the properties of the merging objects, as well as improve the detection and localization of the following GW signal.

Interestingly, December 11th, 2021, a very peculiar gamma-ray burst (GRB) has been detected by \textit{Fermi}/GBM~\citep{Fermi2021}, \textit{Swift}/BAT~\citep{swift2021} and \textit{Insight-}HXMT/HE (GRB211211A)~\citep{insight2021}. An excess in optical/near-infrared has been identified, the multi-band properties of which is quite similar to that of AT2017gfo \citep{Rastinejad2022,xiao2022}. Together with the non-detection of a supernova at the GRB location, this GRB is suggested to be associated with merger event involving a NS, though the duration of the main emission is relatively longer ($\sim8\,$s) compared with typical sGRBs. More intriguingly, a fast rising and exponentially decaying precursor has been observed approximately 1\,s prior to the main emission. The precursor lasts for $\sim0.2\,$s and a quasi-periodic-oscillation (QPO) with frequency $\sim22\,$Hz has been identified in it \citep{xiao2022}. 

Following this interesting observation, various models have been suggested to explain the precursor of GRB211211A. Previous force-free simulations of NS magnetosphere have shown that the interaction of the magnetic fields of two inspiralling-in NSs could produce Poynting flux strong enough to be observed as precursor emissions \citep{carrasco2020}. If a magnetar is involved in the merger event, a catastrophic flare of the magnetar during the inspiral phase could also be the source of the precursor \citep{xiao2022,Zhang2022}. In addition, a resonant shattering of the solid crust of the merging NSs is also invoked to explain the observation, with certain demands on the NS spin and magnetic field \citep{suvorov2022}.

In this paper, we come up with a starquake model to explain the precursor of GRB211211A based on a solid strangeon star scenario, in which the energy budget could be satisfied regardless of the spin and magnetic field strength of the merging NSs. In this model, the equilibrium configuration of the solid compact star changes as the binary gets closer in the inspiral stage and tidal field of the companion becomes stronger. Stress will be accumulated as the elastic structure resists the change in the configuration. Eventually, the elastic strain might exceed a critical value before merger and the solid structure of the star cracks (i.e., a starquake happens). Thus, a precursor will be triggered by energy released during the starquake and the reconfiguration of the star.

The paper will be organized as the follows: in Sec.~\ref{sec:model}, we will introduce the configuration of a solid strange star in the tidal field of its companion; a quantitative comparison with the observation will be made in Sec.~\ref{sec:result}; future observational prospects of this scenario will be discussed in Sec.~\ref{sec:dc}

\section{The model}
\label{sec:model}

\subsection{Equilibrium configuration of a solid strange star in close binary}
The equilibrium configuration of a solid strange star in the tidal field of its companion is determined by the bulk energy of the star $E_\mathrm{total}$, which consists of several parts,
\begin{equation}
E_\mathrm{total}=E_k+E_g+E_t+E_\mathrm{ela},    
\end{equation}
in which $E_k$ is the kinetic energy of the star when rotation is considered, $E_g$ stands for the additional gravitational binding energy of the star, $E_\mathrm{ela}$ is the elastic energy accumulated and $E_t$ is the energy possessed by the star due to the tidal field of its companion.

It is believed that NS spins slowly before merger due to magnetic dipole radiation during the long inspiral stage. Moreover, it has been shown that the tidal interaction is insufficient to synchronize the NS spin before merger~\citep{Bildsten92}. And hence, it is reasonable to assume that both $E_k$ and the change in $E_k$ are negligible before merger. 

$E_g$ and $E_t$ are related to gravity, and it is necessary to define a zero energy configuration as a reference. In the following, we will assume the spherically symmetric configuration to possess energy $E_0$. In binary systems, however, the shape of the star will no longer be spherical and it is useful to define the shape of the star by the parameter of reduced ellipticity 
\begin{equation}
\epsilon=\frac{I-I_0}{I_0}
\end{equation}
in which $I$ is the moment of inertia of the star with arbitrary deformation and $I_0$ is that of the spherical star with the same baryonic mass. For incompressible star and small ellipticity, $\epsilon$
is related to the geometrical eccentricity of the star ($e$) as $\epsilon=e^2/3$. With this definition, one can obtain
\begin{equation}
E_g+E_t=E_0+A_g\epsilon^2-\frac{M_c}{M}A_t(\frac{R}{D})^3\epsilon,
\end{equation}
in which $M$ and $R$ are the mass and radius of the star, $M_c$ is the mass of its companion, $D$ is the separation of the binary system, $A_t$ and $A_g$ are coefficients related to gravitational binding energy which depend on the density distribution of the star. In the case of incompressible star (which is a good approximation for strange stars), the following relation holds 
\begin{equation}
A_g=\frac{3}{25}A_t=\frac{3}{25}\frac{GM^2}{R}.
\end{equation}
For simplicity, we will use the notation $A=GM^2/R$ in the expressions below and $A\sim1\times10^{54}\,$erg for typical NS value $M=1.4\,M_\odot$ and $R=10\,$km.

The elastic energy $E_\mathrm{ela}$ is related to difference between the reference ellipticity $\epsilon_0$ when the star solidified (for example, when the star suffers the previous starquake) and the current value $\epsilon$. According to Hooke's Law, one can obtain
\begin{equation}
E_\mathrm{ela}=B(\epsilon-\epsilon_0)^2
\end{equation}
in which $B=\frac{1}{2}\mu V$, $\mu$ and $V$ are the shear modulus and volume of the solid strange star, respectively. 

Collecting all the ingredients we have 
\begin{equation}
E_\mathrm{total}=E_0+\frac{3}{25}A\epsilon^2-A\frac{M_c}{M}(\frac{R}{D})^3\epsilon+B(\epsilon-\epsilon_0)^2
\end{equation}
and the ellipticity of equilibrium configuration could be obtained by minimizing the total energy which requires $\partial E_\mathrm{total}/\partial\epsilon=0$. Thus, we could obtain the reduced ellipticity 
\begin{equation}
\epsilon=\frac{25A}{6A+50B}\frac{M_c}{M}(\frac{R}{D})^3+\frac{50B}{6A+50B}\epsilon_0.
\label{eq:epeq}
\end{equation}
For a purely fluid star (or for a solid star when starquake happens, to reconfigurate itself), the equilibrium ellipticity is then
\begin{equation}
\epsilon_\mathrm{eq,fl}=\frac{25}{6}\frac{M_c}{M}(\frac{R}{D})^3.
\label{eq:epeqfl}
\end{equation}
It is worth noting that the solid star tends to resist from being deformed due to the accumulation of elastic energy. And hence, if the star were at its fluid equilibrium configuration before the previous solidification, then at any time during the later evolution, we should have $\epsilon<\epsilon_\mathrm{eq,fl}$. Additionally, the tidal interaction becomes stronger as the binary gets closer during the inspiral stage, the ellipticity of the star would increase with time, so we could obtain the following inequality which holds during the entire inspiral stage
\begin{equation}
\epsilon_0<\epsilon<\epsilon_\mathrm{eq,fl}.
\label{eq:epineq}
\end{equation}

Previous calculation and observations indicate that the shear modulus of a strange star $\mu$ lies in the range between $10^{30}$ and $10^{34}\,\mathrm{erg/cm^3}$~\citep{Xu2003,zhou2004}. Within this range, the value of $B$ would be much smaller than $A$. Combining the inequality Eq.~\eqref{eq:epineq}, one could obtain that the second term in Eq.~\eqref{eq:epeq} is negligible compared with the first term. Therefore, we will omit the second term in Eq.~\eqref{eq:epeq} in the calculations below and assume the equilibrium configuration of a solid strange star satisfies
\begin{equation}
\epsilon_\mathrm{eq,so}=\frac{25A}{6A+50B}\frac{M_c}{M}(\frac{R}{D})^3.
\label{eq:epeqso}
\end{equation}

\subsection{The starquake model}

Eq.~\eqref{eq:epeqso} indicates how the ellipticity of the solid strange star increases as the separation of the binary shrinks during the inspiral stage, due to the dissipation of angular momentum and energy through GW radiation. Elastic energy gradually increases as the shape of the solid star changes. Depending on the microscopical model of the solid strange star, the maximum stress of the solid structure might be reached before the merger happens and hence a starquake takes place. During the starquake, the ellipticity of star tends to migrate from its solid equilibrium configuration ($\epsilon_\mathrm{eq,so}$) to its fluid case ($\epsilon_\mathrm{eq,fl}$). Therefore, not only the elastic energy but also 
the change in $E_g+E_t$ will be released during the starquake as they are related to the ellipticity. 
This starquake scenario is quite similar to the starquake model of pulsar glitches~\citep{Baym71c,zhou2004,zhou2014,Lai2018c}. In the latter case, the accumulation of the elastic energy (i.e., the change in the ellipticity) is due to the change of pulsar spin, which results from the magnetic dipole radiation of pulsars. 

We will first calculate the change in $E_g+E_t$  before and after the starquake. The change of the reduced ellipticity is the difference between $\epsilon_\mathrm{eq,fl}$ and $\epsilon_\mathrm{eq,so}$:
\begin{equation}
\delta\epsilon=\epsilon_\mathrm{eq,fl}-\epsilon_\mathrm{eq,so}=\frac{50B}{6A+50B}\frac{25}{6}\frac{M_c}{M}(\frac{R}{D})^3.
\end{equation}
According to the feasible range of $\mu$, $B$ is at least two orders of magnitude smaller than $A$ and hence we have $\delta\epsilon<0.1\epsilon_\mathrm{eq,fl}$. Therefore, we could estimate the 
change in $E_g+E_t$ by an expansion with respect to $\epsilon$:
\begin{equation}
\delta(E_g+E_t)
\sim\frac{\partial(E_g+E_t)}{\partial\epsilon}|_{\epsilon_\mathrm{eq,fl}}\delta\epsilon.
\label{eq:deltaeget}
\end{equation}
However, $\epsilon_\mathrm{eq,fl}$ is obtained by requiring $\partial(E_g+E_t)/\partial{\epsilon}=0$. Consequently, the non-vanishing parts of $E_g+E_t$ are terms of $\delta\epsilon^2$ and higher order ones and hence are negligible. Similar conclusion is found for the case of starquake scenario of pulsar glitches~\citep{zhou2014}. As a result, the major contribution of energy release during such starquake is the release of the accumulated elastic energy before the starquake:
\begin{equation}
E_\mathrm{ela}=B(\epsilon_\mathrm{eq,so}-\epsilon_0)^2=B[\frac{25A}{6A+50B}\frac{M_c}{M}(\frac{R}{D})^3-\frac{25}{6}\frac{M_c}{M}(\frac{R}{D_0})^3]^2,
\end{equation}
in which $D_0$ is the separation of the binary system when the solid strange star experienced its previous solidification. As the precursor happened very close to the merger (i.e., when the orbital separation shrinks rapidly), it is reasonable to make the following assumption $D_0\gg D$. In addition, we remind that $A\gg B$, then the elastic energy is approximately
\begin{equation}
E_\mathrm{ela}\sim B [\frac{25}{6}\frac{M_c}{M}(\frac{R}{D})^3]^2.
\label{eq:eela}
\end{equation}
It is easy to verify that the total change in $E_g+E_t$ is indeed $\sim10\%$ of $E_\mathrm{ela}$ at most and we will focus on Eq.~\eqref{eq:eela} when comparing with the observation in the next section.

\section{Results}
\label{sec:result}
\subsection{Energy budget}
The origin of GRB211211A is still under debate. Scenarios including binary neutron star (BNS) merger~\citep{Kunert2023}, black hole-neutron star (BH-NS) merger~\citep{Zhu2022} as well as neutron star-white dwarf (NS-WD) merger~\citep{yang2022} have been proposed. In this section, we will focus on BNS and BH-NS merger scenarios and test their feasibility and parameter space in producing the observed precursor within the starquake scenario. 

The major difference between BNS and BH-NS merger scenarios is their mass ratio $M_c/M$ and the orbital separation $D$ when the starquake happens. For BNS scenario, the most mass-asymmetric BNS system observed and the mass of which is precisely measured has mass ratio $q=0.78$~\citep{Ferdman2020}. GW190425 was measured to have an even smaller mass ratio of $0.7$, whereas it is still uncertain that whether its heavier component is a NS or a BH~\citep{Abbott2020}. Even though there are equation of state (EoS) models for solid strange stars which could reach maximum mass as high as above 3$\,M_\odot$~\citep{lai2009}, this could merely push the possible range of $q$ to about 0.5 (or, $M_c/M\sim2.0$ when we consider the lighter star suffers the starquake). 

For BH-NS merger case, the range of the binary mass ratio could be much wider. GW observations indicate that $M_\mathrm{BH}/M_\mathrm{NS}$ could be as small as 2.0, if heavier components of those mass gap events are indeed BHs. The upper limit for $M_\mathrm{BH}/M_\mathrm{NS}$ for the event GRB211211A should be constrained by the observation of associated kilonova. If the mass of the BH is too much larger than NS, the ejected mass from the NS would be insufficient for powering a kilonova even with an extreme BH spin. According to previous studies, we put 10.0 as a maximum possible value for $M_\mathrm{BH}/M_\mathrm{NS}$~\citep{Kyutoku2015,Kawaguchi2016}. In the analysis below, we will assume $M_c/M$ to be in the range of $[2.0,10.0]$ for BH-NS merger scenario.

\begin{figure}[h]
\centering
\includegraphics[width=0.5\textwidth]{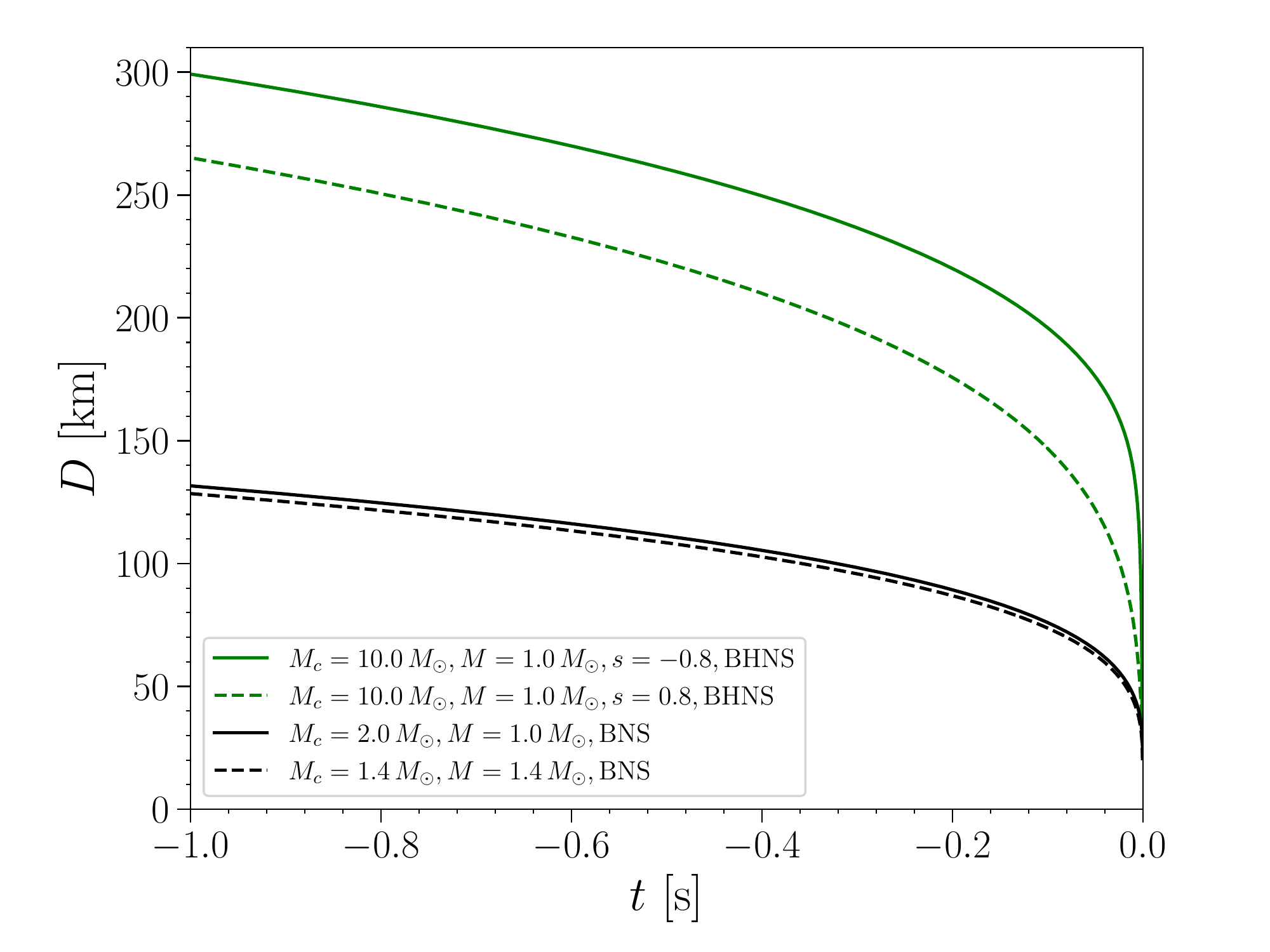}
\caption{The evolution of the orbital separation $D$ prior to the merger (the merger time is calibrated to $t=0$ in this figure) for BNS and BH-NS cases, obtained by effective-one-body (EOB) calculations~\citep{Riemenschneider2021,Nagar2020,Nagar2019,Nagar2018,Nagar2015,Damour2014}. The green solid and dashed curves on the top represent the BH-NS merger scenario with $M_c/M=10$ and BH spin of $-0.8$ (i.e., anti-aligned with the orbital angular momentum) and $0.8$, respectively. The black solid and dashed curves on the bottom stand for the BNS merger cases with $M_c/M=1.0$ and $2.0$, respectively.}.
\label{fig:dvst}
\end{figure}

The orbital separation $D$ when the starquake happens could be implied by the time when the precursor happens. For GRB211211A, the time between the precursor and the main burst is approximately one second, which means the starquake happens less than one second before merger happens as there might be time delay between the actual merger and the time when jet is launched (for instance, for the BNS merger GW170817, the time delay between the merger and sGRB is approximately 1.7 seconds). For BNS system this indicates an orbital separation $D<\sim100\,$km according to Fig.~\ref{fig:dvst}. For BH-NS system the possible range of $D$ is larger (i.e., could be as large as $\sim300\,$km), due not only to a wider range of the mass range, but also to BH spin which could affect the dynamics of the final inspiral phase. Particularly, extreme spin is necessary for significant mass ejection to take place for BH-NS system with large mass ratio.

For BNS case, inserting the value of 50\,km for $D$, one can obtain the total elastic energy contained in the entire star as the following
\begin{equation}
E_\mathrm{ela}=2.3\times10^{49}\,\mathrm{erg}(\frac{\mu}{10^{34}\,\mathrm{erg\,cm^{-3}}})(\frac{M_c/M}{1.0})^2(\frac{D}{50\,\mathrm{km}})^{-6}.
\label{eq:eelanum}
\end{equation}
For the observation of GRB211211A, the energy released in the precursor in the form of electromagnetic emission is $\sim7.7\times10^{48}\,\mathrm{erg}$ which means the total energy released should be larger than this value. Therefore, the energy budget is very tense for explaining the precursor in the BNS case, unless the solid structure of the entire star shatters and the starquake has to happen at a closer range with the jet launching being delayed after the merger. Alternatively, one can count on a very mass-asymmetric merger with $M_c/M$ as large as 2.0. Nevertheless, even in this case, the total elastic energy is not enough if $D$ is larger than 100\,km when the starquake happens. 

\begin{figure}[h]
\centering
\includegraphics[width=0.5\textwidth]{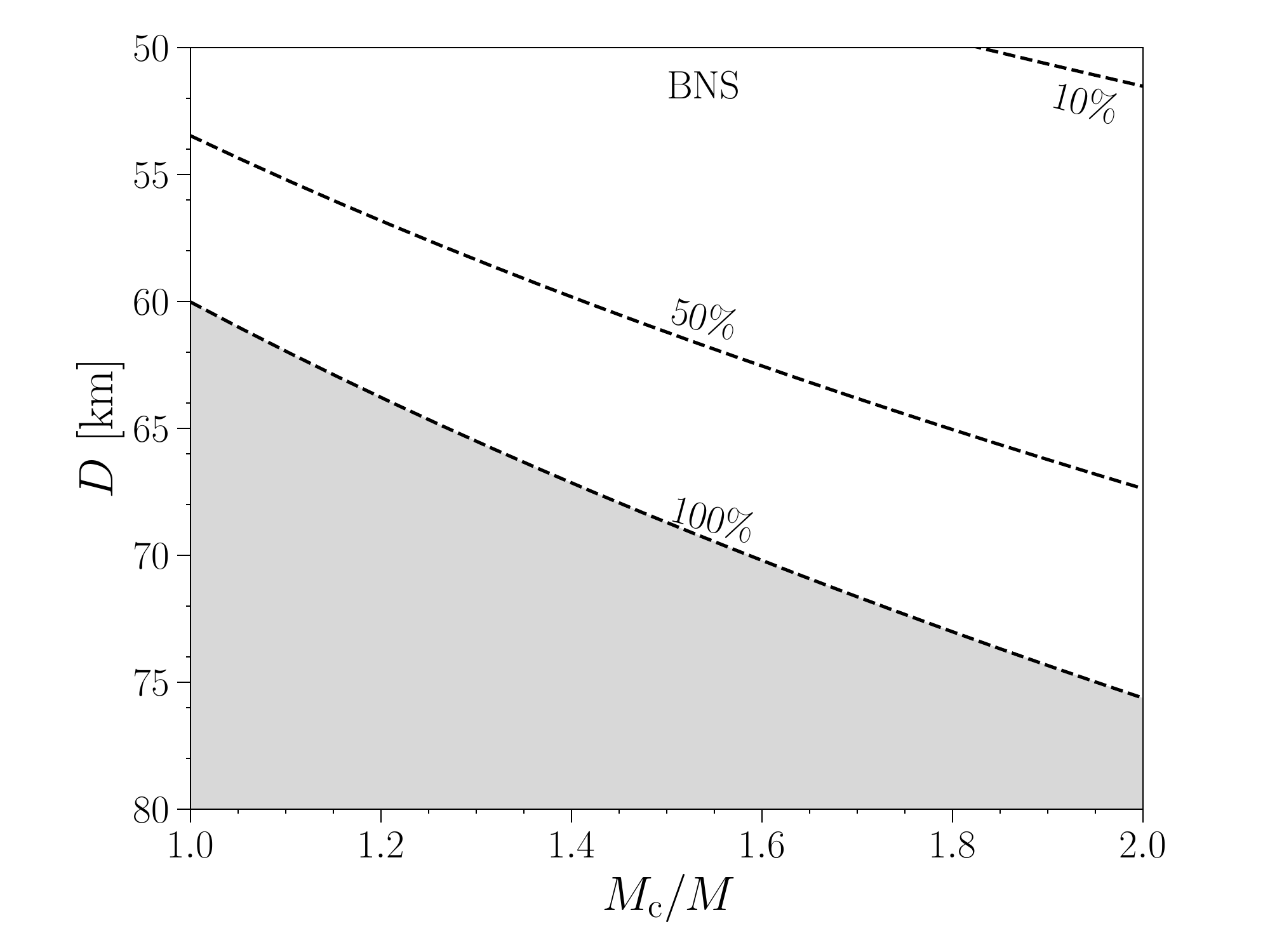}
\caption{Constraints on the binary parameters according to the precursor observation of GRB211211A in the BNS case. Three dashed lines labeled by $100\%$, $50\%$ and $10\%$ represent the combination of the mass ratio $M_c/M$ and orbital separation $D$ when the starquake happens with which the corresponding percentage of the total elastic energy has to be released in order to explain the energy of the precursor observation. The grey shaded region is excluded since more than $100\%$ of the total energy budget is required (and hence insufficient). If we consider that partial failure of the solid structure (i.e., less than $10\%$ of the elastic energy is released) is more realistic, the possible paremeter space is very narrow unless the starquake happens extremely close to the merger ($D$ smaller than 50\,km).}
\label{fig:nsns}
\end{figure}

The detailed result is shown in Fig.~\ref{fig:nsns}. Even if we allow for a global starquake during which the elastic energy of the entire star is released and all converted into precursor EM emissions, the orbital separation at the moment of this starquake has to be smaller than $\sim60\,$km for an equal mass BNS merger. This upper limit on $D$ could be relaxed to $\sim75\,$km if we consider extreme mass ratio cases. Nevertheless, as mentioned above, such global starquake is less likely to happen and a partial starquake which releases $10\%$ of the elastic energy contained in the star is not possible unless the starquake happens at a binary separation less than $50\,$km and with large mass ratio.   

Compared with the BNS case, BH-NS merger scenario is more favored considering the energy budget due to a larger possible $M_c/M$. Ten percents of the total elastic energy converting into EM emission would be enough for the observed precursor luminosity. According to previous researches, it is shown that elastic energy is not uniformly accumulated throughout the solid star when deformation is induced. Partial failure in the solid structure is also demonstrated to be more reasonable to explain the observation of pulsar glitches~\citep{Lai2018c}. In our case, the deformation is induced by the tidal field of the companion of a solid strange star. Consequently, the stress is expected to be the largest near the surface of the star, especially near the polar and equatorial region. Thus a small fraction of the solid star shatters and releases the elastic energy in this region is a more natural scenario. Indeed, there would be larger uncertainties in inferring the orbital separation by the time before merger for BH-NS case. Therefore, we treat $D$ as a free parameter and explore the possible combination for $D$ and $M_c/M$ for different amount of elastic energy required.

\begin{figure}[h]
\centering
\includegraphics[width=0.5\textwidth]{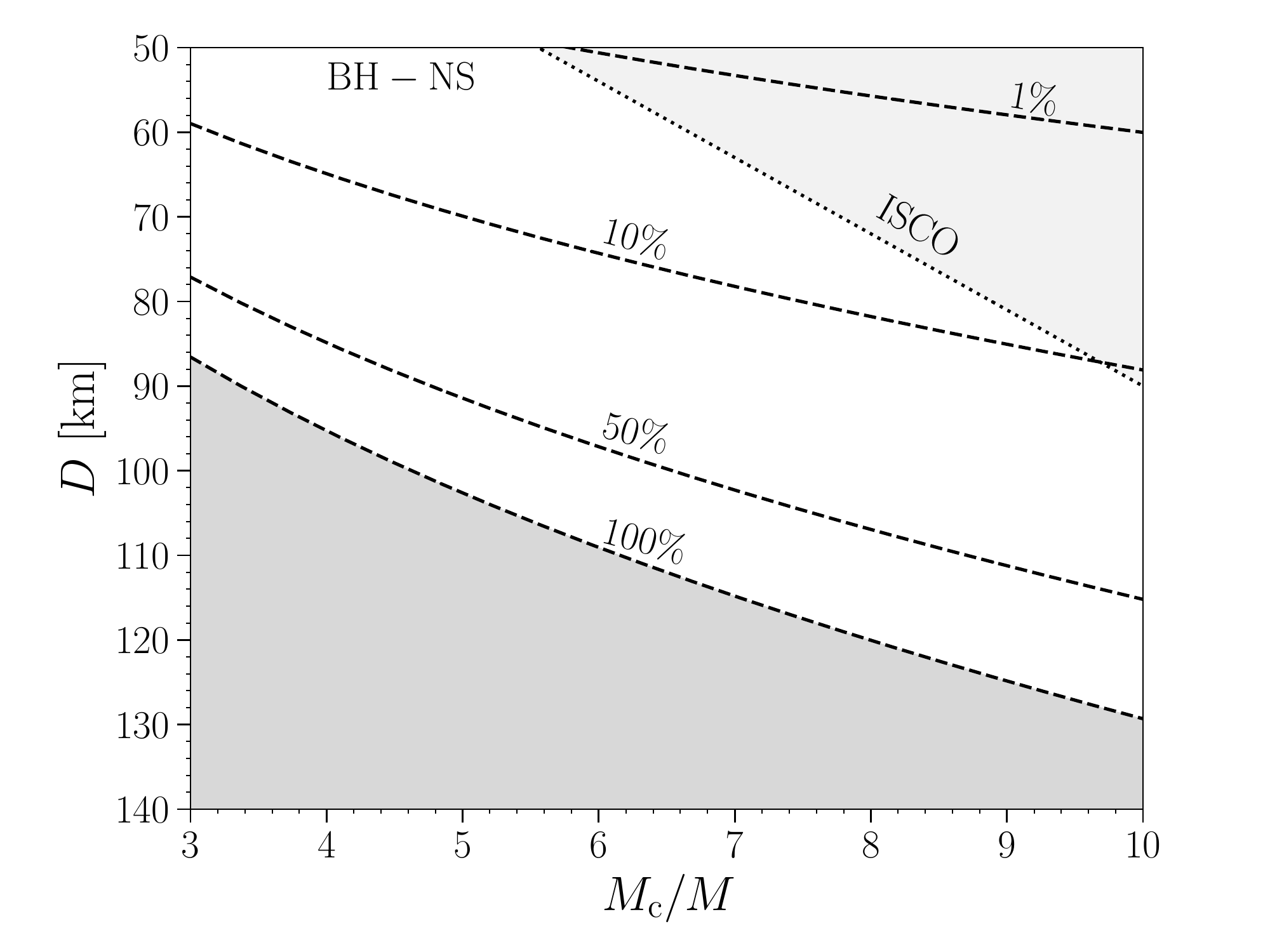}
\caption{Constraints on the binary parameters according to the precursor observation of GRB211211A in the BHNS case. Three dashed lines are plotted in the same way as in Fig.~\ref{fig:nsns}. The darker bottom left grey shaded region is excluded according to the energy budget whereas the lighter grey shaded region on the top right is inside the ISCO of the BH companion (assuming that the mass of the NS is $1\,M_\odot$ and the BH is a Schwarzschild BH) and hence is excluded. Nevertheless, the ISCO of a spinning BH can be much smaller than that of a Schwarzschild BH and parameters in this region could as well be possible, as extreme spin is essential for sufficient mass ejection in the case of large mass ratio BH-NS mergers. }.
\label{fig:bhns}
\end{figure}

The results for the allowed parameter space of the BH-NS case are shown in Fig.~\ref{fig:bhns}. As the larger allowed mass ratio benefits the accumulation of elastic energy, the separation of the binary could be as large as $\sim100\,$km when the starquake happens if we allow for a $100\%$ release and conversion of the elastic energy. There is still plenty of possible parameter space even if we consider a partial ($10\%$) elastic energy release for orbital separation $D\sim60\,$km. For even smaller fraction (such as $1\%$ case), the binary separation at which the starquake happens lies inside the innermost stable circular orbit (ISCO) of a Schwarzschild BH with the same mass of the BH companion. However, as mentioned above, large spin is needed for large mass ratio ($M_c/M\sim$5) BH-NS merger to produce a kilonova and the ISCO radius of a Kerr BH could be much smaller than that of the Schwarzschild case. Consequently, it is still possible for a tiny starquake to happen at such small separation and account for the precursor observation of GRB211211A.

\subsection{Sequential starquakes during the insipiral}
\label{subsec:sequential}

\begin{figure}[h]
\centering
\includegraphics[width=0.5\textwidth]{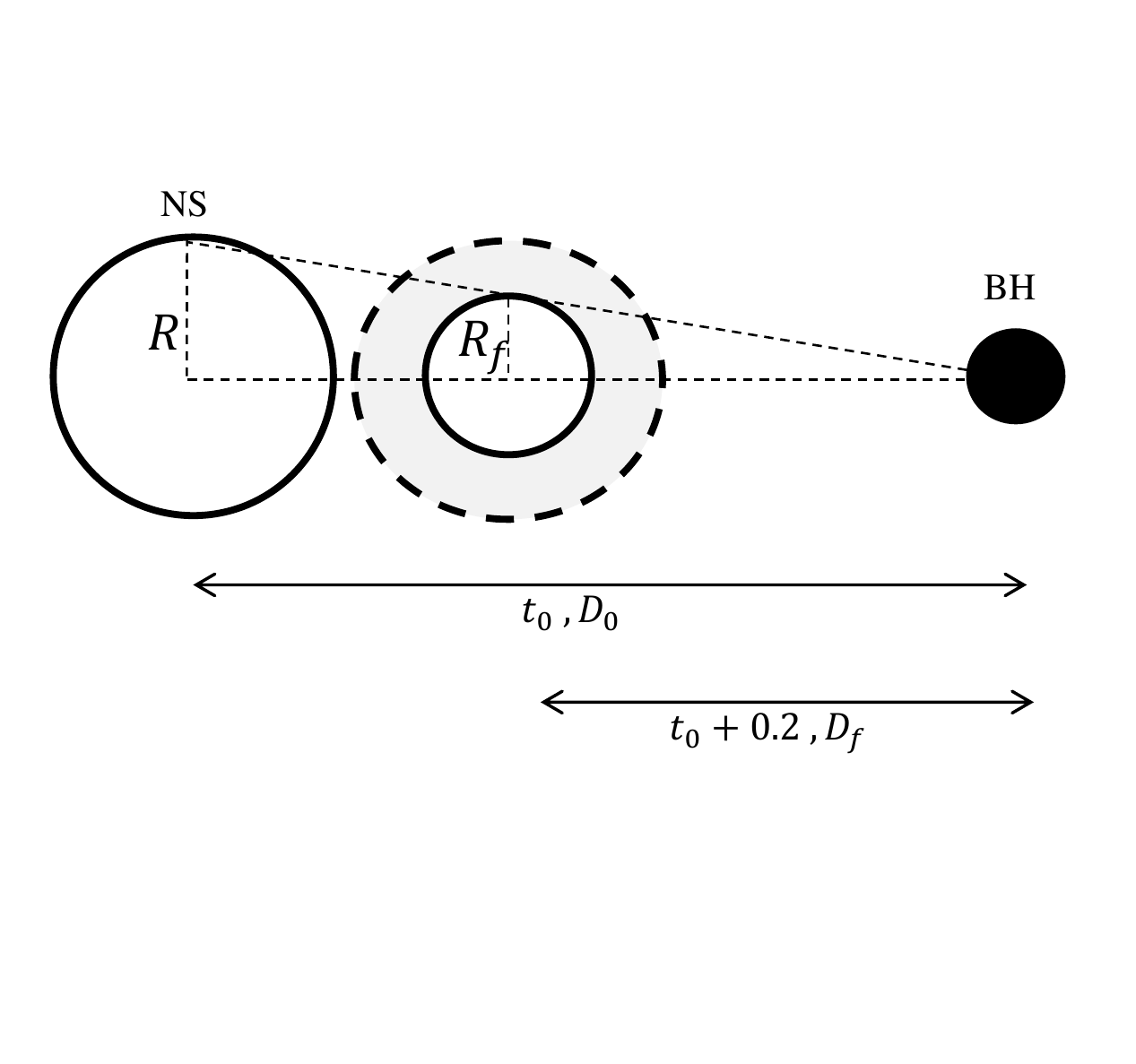}
\caption{An illustration of the sequential starquake model as discussed in Sec.~\ref{subsec:sequential}. The first starquake happens at $t_0$ with binary orbital separation $D_0$. After that, sequential starquakes take place with increasing depth inside the solid strange star, as the binary gets closer. At a later time (for instance, $t_0+0.2\,$s), only the matter inside a sphere with radius $R_f$ could maintain its solid structure, with $R_f$ approximately satisfies the following relation: $R_f/D_f=R/D_0$. Grey shaded region illustrates the part of star of for which the elastic energy could be released during this sequential starquake. }.
\label{fig:illu}
\end{figure}

The above discussion is based on the assumption that the deformation induced is homogeneous in the entire star and only one starquake happens prior to the merger, which could either be a global quake or a local one. Nevertheless, the later in the inspiral stage, the faster the binary orbit shrinks (cf. Fig.~\ref{fig:dvst}) and hence the elastic energy is accumulated much more rapidly. Moreover, the ellipticity induced by the tidal field increases as $(R/D)^3$ which suggests the deformation of the star is larger in its outer part for a realistic star (i.e., which could not be incompressible) at a fixed separation. With these considerations, it is more likely that several sequential starquakes take place during the last of the inspiral phase, if the first starquake is a partial one. However, in this case elastic energy is partially released and we focus on the BH-NS scenario in this section, due to its larger parameter space.

Our sequential starquake model is demonstrated in Fig.~\ref{fig:illu}. The first starquake takes place at a separation of $D_0$ for the binary at time $t_0$. As the deformation is largest in the outer part, only the surface part of the star suffers the quake (for instance, the region from $R-\delta R$ to $R$, in which $\delta R$ is the depth of this starquake). At a later time $t_f$, the binary would be separated at a smaller distance of $D_f$, and a shell of the star at a radius of $R_f$ would experience the same deformation as those at the surface of the star at the separation of $D_0$, in which $R_f$ is simply determined by the geometrical relation $R_f/D_f=R/D_0$. Thus, the matter at $R_f$ meets its limit for a failure in its elastic structure. In this case, several individual starquakes take place during the time $t_0$ to $t_f$, from the surface of the star to the depth of $R-R_f$. 

The observations of the precursor of GRB211211A indicate that the energy is released in several individual bursts with decaying peak amplitude in a time span of $\sim0.2\,$s, which is also the origin of the 22\,Hz QPO. Such feature could be well explained by a sequential starquake model: those later starquakes happen deeper inside the star and consequently the energy release tends to be diffused more slowly and hence the peak flux observed would be lower. 

In such a model, the fraction of elastic energy released during the sequential starquakes over a certain duration depends on the binary separation of the first/last quake. As a result, in the sequential starquakes scenario, there will be additional constraint on the parameter space. The fraction of the elastic energy released in $0.2\,$s could be obtained as 
\begin{equation}
\frac{E_\mathrm{release}}{E_\mathrm{ela}}=\frac{V_\mathrm{quake}}{V_\mathrm{total}}
=1-(\frac{D_f}{D_0})^3,
\end{equation}
in which $E_\mathrm{ela}$ also depends on $D_f$ as in Eq.~\ref{eq:eelanum}. Therefore, given the mass ratio and BH spin, $D_0$ could be determined by the EOB model and hence $E_\mathrm{release}$ could be obtained. The fact that the energy released has to be greater than the observed value then sets a constraint on $D_f$.

\begin{figure}[h]
\centering
\includegraphics[width=0.5\textwidth]{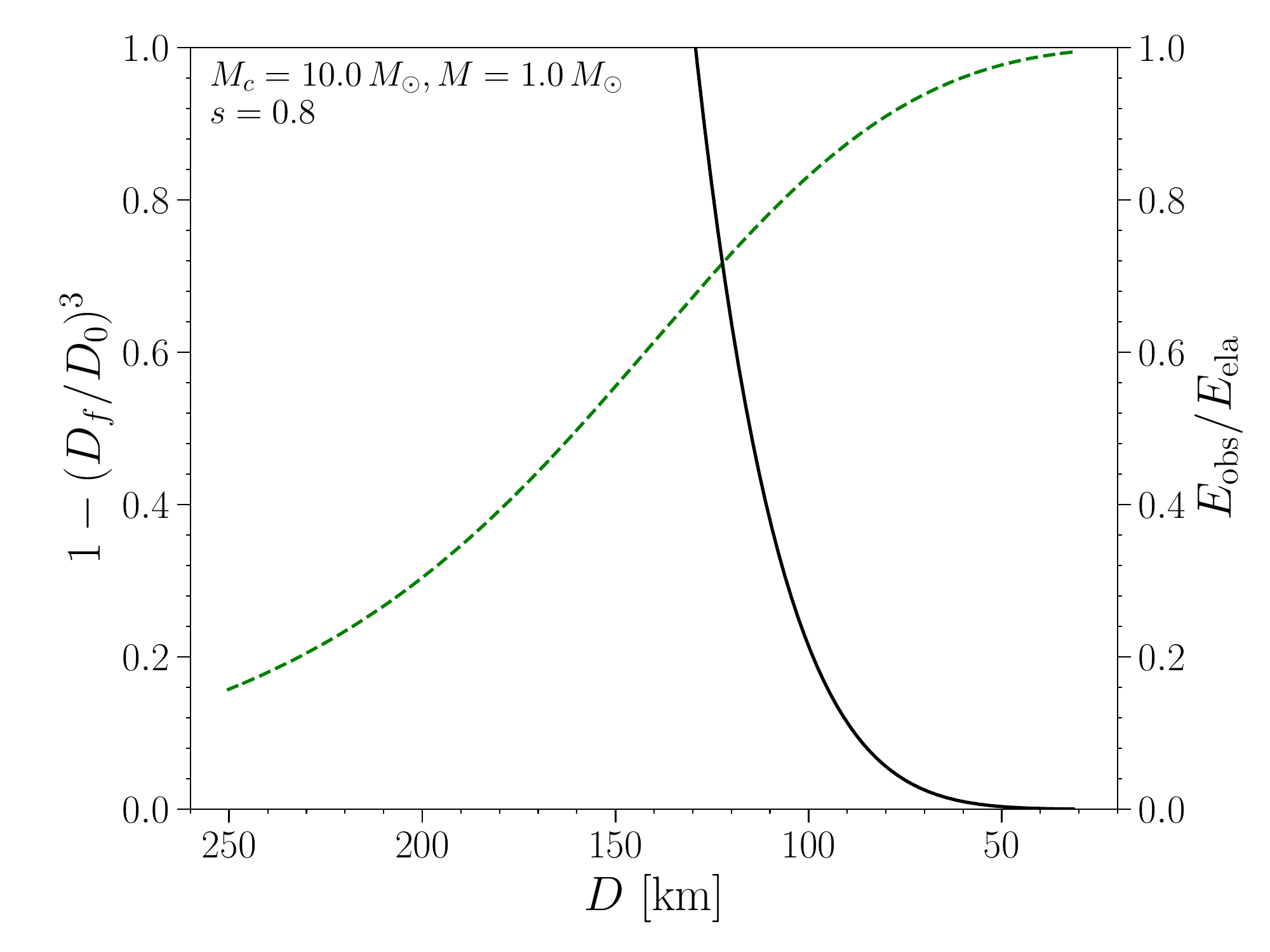}
\caption{Constraint on the binary separation when the last quake takes place in the sequential quakes scenario. The mass ratio and BH spin is $M_c/M=10$ and $s=0.8$ for this model. With the green dashed curve we plotted the fraction of the volume of the star which suffers starquakes in the duration of $0.2\,$s ($1-(D_f/D_0)^3$), which is equal to the fraction of elastic energy released, assuming a uniform shear modulus. The black curve shows the ratio between the observed precursor energy and the total elastic energy ($E_\mathrm{obs}/E_\mathrm{ela}$). The intersection point then sets up an upper limit for the parameter $D_f$. }.
\label{fig:resq10}
\end{figure}

\begin{figure}[h]
\centering
\includegraphics[width=0.5\textwidth]{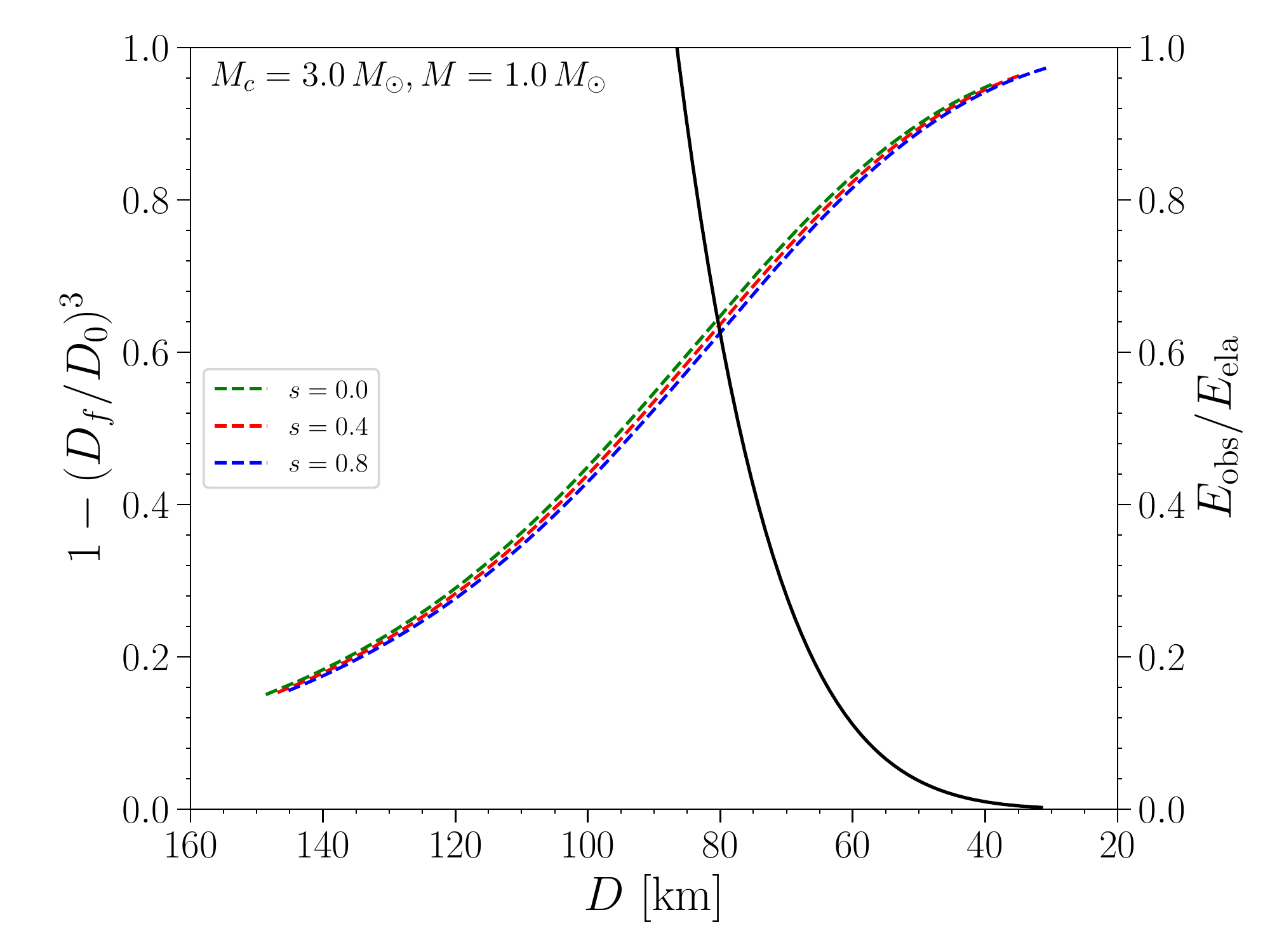}
\caption{The same as Fig.~\ref{fig:resq10} but for a mass ratio of $M_c/M=3$. In this case we have also explored different values of BH spin to check its influence on our analysis, the green dashed curve, red dashed curve and blue dashed curve stand for the cases of $s=0,\,0.4$ and 0.8. }.
\label{fig:resq3}
\end{figure}

The result is shown in Fig.~\ref{fig:resq10} and Fig.~\ref{fig:resq3}, in which we have chosen the case of $M_c/M=10$ and 3, respectively. As mentioned above, for larger mass ratio case, large BH spin is also essential \citep{Kyutoku2015,Kawaguchi2016} and we have chosen $s=0.8$ for the dimensionless spin for the $M_c/M=10$ case. For smaller mass ratio, we have explored 3 different BH spin ($s=0,\,0.4$ and 0.8) parameter to verify its impact on our results. Our result shows that constraint is not very sensitive on the BH spin and upper limit for $D_f$ is approximately 120\,km for $M_c/M=10$ and 75\,km for $M_c/M=3$, which are all much larger than the ISCO radius. It is worth noting that, if one assumes a certain energy conversion efficiency ($\eta$) from the released elastic energy to the EM emission, a tighter constraint could be made by requiring
\begin{equation}
\eta=\frac{E_\mathrm{obs}/E_\mathrm{ela}}{1-(D_f/D_0)^3}
\end{equation}
which could be read from the result figures once a value of $\eta$ is assumed.

In this model, the rising timescale of the precursor could also be understood. The change in the ellipticity of the star after the first quake happens approximately in a free falling timescale as 
\begin{equation}
t_\mathrm{rise}\sim\frac{2\pi R}{\sqrt{GM_c/D_0}}.
\end{equation}
The timescale is larger for smaller mass ratio case as $M_c/D_0$ becomes smaller according to the analysis above. In the $M_c/M=3.0$ case, the above formula yields a rising timescale of $\sim0.8\,$ms, which is consistent with the observations.

In addition, the QPO frequency in the observation of the precursor of GRB211211A could also be naturally understood as the number of quakes during the entire process within 0.2\,s. However, the number of quakes and the energy released in each quake in the entire sequential process depend on the thickness of the shells which suffer the failure of the solid structure during each quake. The thickness of the shattered shells could be very sensitive to properties which are difficult to predict such as crystal defects in the solid structure which might be caused during previous quakes. And hence, the observational properties (i.e., the QPO frequencies) could be totally different even for sources with similar binary parameters.

\subsection{Quake-induced Oscillation?}

The enormous energy released during a tide-induced giant quake may set the entire star into vibration, producing oscillations at frequencies determined by the structure and elastic properties of the solid strange star. Typically, non-radial oscillations of solid stars include spheroidal and toroidal modes. Toroidal (torsional) modes are a type of oscillation that maintains the star's shape. They are purely shear oscillations. Quasi-periodic oscillations (QPOs) following giant flares in soft gamma-ray repeaters and anomalous X-ray pulsars indicate the shear motion of the star after a giant quake~\citep{Duncan:1998my,Watts:2006mr}. The spheroidal oscillations refer to waves that change the shape of the star, involving both radial and tangential displacements. These can be easily excited in a tidally-induced starquake. Both types of oscillations have been observed in large earthquakes~\citep{Benioff1961,Park2004}. Furthermore, these modes couple more easily with external magnetic fields than modes coming from the deep interior of the star.
For these reasons, we expect that the tide-induced torsional modes may couple with Alfvén waves along the magnetospheric field lines, resulting in the QPO in the precursor of GRB211211A. 

We denote $(\xi_{r}, \xi_{\theta}, \xi_{\phi})$ as the displacement amplitudes in spherical coordinate. To estimate the frequencies of toroidal and spheroidal modes, we model the solid strange star as a homogeneous, isotropic elastic sphere with uniform density and shear modulus.

The toroidal oscillations are divergence free with no radial components.
For a particular eigenmode denoted by $l$ and $m$, the separation of variable for the displacements takes the form~\citep{McDermott1988} 
\begin{equation}
	\label{eqn:variables}
	\xi_\theta= \frac{W(r)}{\sin \theta} \frac{\partial Y_{l m}}{\partial \phi} e^{i \omega t}, 
	\quad 
	\xi_\phi=-W(r) \frac{\partial Y_{l m}}{\partial \theta} e^{i \omega t}\,.
\end{equation}
where $Y_{lm}(\theta,\phi)$ are spherical harmonics. 
Inserting Eq.~(\ref{eqn:variables}) into the shear wave equation, we obtain the following equation for the radial eigenfunction $W(r)$~\citep{McDermott1988} 
\begin{equation}
	\frac{{\rm d}^2 W}{{\rm d} r^2}+\frac{2}{r} \frac{{\rm d} W}{{\rm d} r}+\left[\frac{\rho \omega^2}{\mu}-\frac{l(l+1)}{r^2}\right] W=0\,.
\end{equation}
where $\rho$ and $\mu$ are the star's density and shear modulus respectively. The torsional modes are referred by the notation $_n t_{l}$, where $n$ is the overtone number of radial nodes in the eigenfunction $W(r)$. In this paper, we focus on the nodeless vibrations with $n=0$. By applying the boundary condition of vanishing surface horizontal traction, we can solve the eigenvalue problem analytically~\citep{Lamb1881}, obtaining the eigenfrequency
\begin{equation}
        \label{eqn:torsional}
	f \left(_0 t_{l}\right)= \mathcal{C}_{l}\frac{v}{ R}\,.
\end{equation}
Here $v=(\mu/\rho)^{1/2}$ is the shear speed, $\mathcal{C}_{l}$ is a constant depending on $l$. The fundamental mode of torsional oscillation is $_0 t_2$ because $_0t_{0}$ has zero displacement and $_0 t_1$ describes a constant azimuthal twist of the entire star, which does not exist for free oscillation.

We denote the spheroidal mode as $_0 s_{l}$. The separation of variable for the displacement can be written as 
\begin{equation}
\xi_r=U(r) Y_{l m}, \quad \xi_\theta=V(r) \frac{\partial Y_{l m}}{\partial \theta}, \quad \xi_\phi=\frac{V(r)}{\sin \theta} \frac{\partial Y_{l m}}{\partial \phi}\,,
\end{equation}
Combined with perturbation in the gravitational potential, we can obtain the systematic differential equations governing the oscillation. We refer the readers to Refs.~\citep{McDermott1988,Crossley1975,Alterman1959} for those equations. To simplify the calculations and represent the eigenfreqeuency in a simple relation to the shear speed, we use the long wavelength approximation. The eigenfrequency of nodeless modes is 
\begin{equation}
\label{eqn:spheroidal}
    f\left({ }_0 s_{l}\right)=\frac{[2(2 \ell+1)(\ell-1)]^{1 / 2}}{2\pi}\frac{v}{R}\,.
\end{equation}
The oscillation of order $l=1$ doesn't exists, and the fundamental mode of the global nodeless spheroidal oscillation is $_0s_2$.

Both the strong and the electromagnetic interactions are involved in characterizing the shear modulus, $\mu$, of solid strange matter. One may have $\mu\sim 10^{34}\,\rm erg\,cm^{-3}$ if the strong force dominates the shear mode~\cite{Xu2003}. In the catastrophic process of the tide-induced giant quake, the star may be fractured as a whole and release enormous energy. The modulus could be decreased after the fracture.

\begin{figure}
\centering
\includegraphics[width=0.5\textwidth]{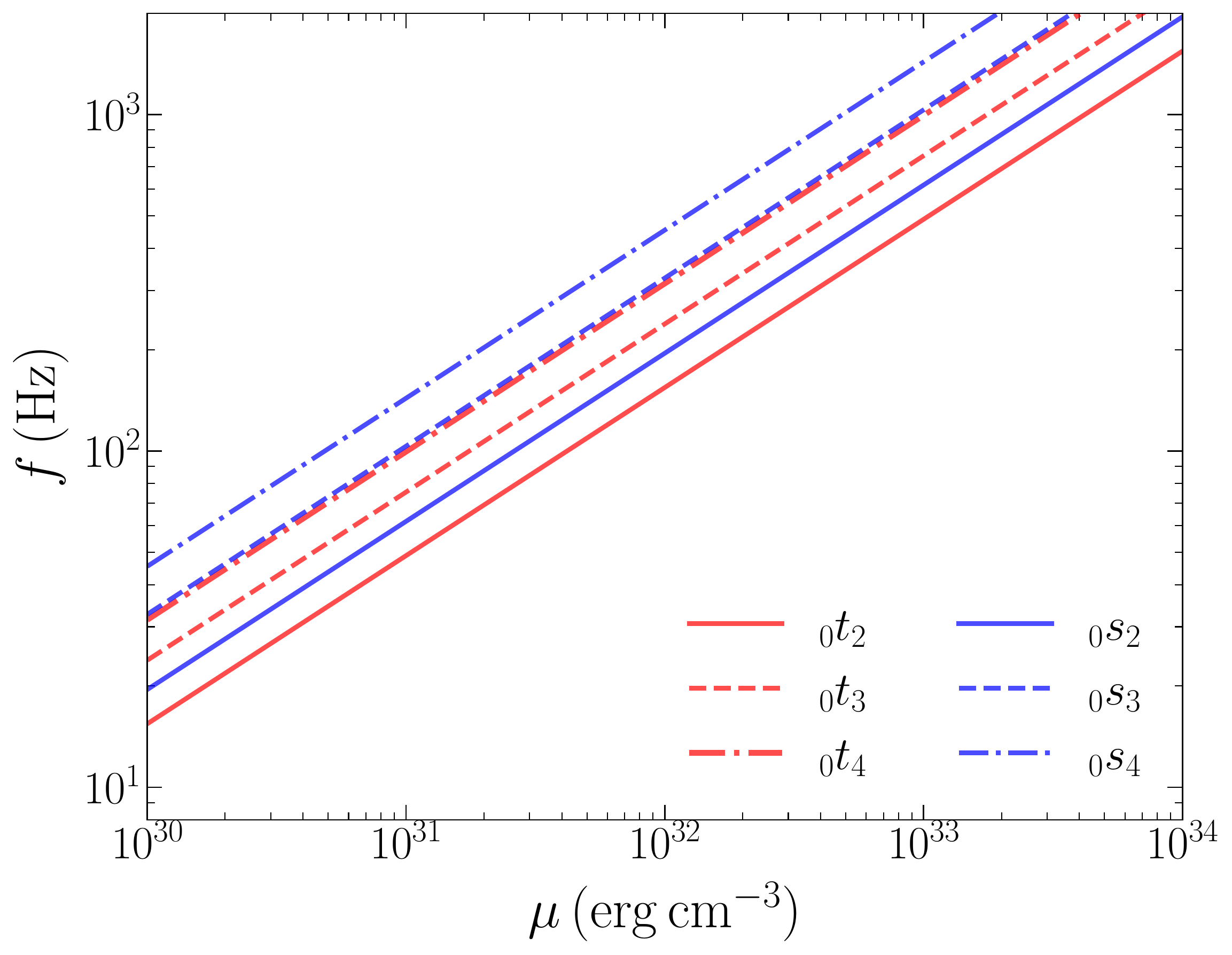}
\caption{The relation between eigenfrequencies of the fundamental modes of the torsional and the spheroidal oscillations. We show the cases with $l=2, 3, 4$.}.
\label{fig:oscillation}
\end{figure}

From Eq.~(\ref{eqn:torsional}) and Eq.~(\ref{eqn:spheroidal}), we obtain the fundamental frequencies for the torsional and spheroidal modes 
\begin{align}
    f(_0t_{2})=&48.8 \,{\rm Hz}\left(\frac{\mu}{10^{31}\,\rm erg\,\rm cm^{-3}} \right)^{\frac{1}{2}}\,,\\
    f(_0s_{2})=&61.7 \,{\rm Hz}\left(\frac{\mu}{10^{31}\,\rm erg\,\rm cm^{-3}} \right)^{\frac{1}{2}}\,.
\end{align}
Here we take $M=1.4\,M_{\odot}$ and $R=10\,\rm km$. One can notice that the mode frequency is on the same order of the QPO frequency observed in GRB211211A if the shear modulus of the solid strange star decreases to the order of  $10^{30}-10^{31}\,\rm erg\,\rm cm^{-3}$. In Fig.~\ref{fig:oscillation}, we show the fundamental frequencies with $l=2, 3, 4$ as functions of the shear modulus.

In reality, the frequency of a mode will be shifted due to the effects of gravitational redshift and the Doppler effect. The gravitational redshift factor caused by the NS can be expressed as
\begin{equation}
z_{\rm NS}=\left( 1-\frac{2GM}{Rc^2}\right)^{-\frac{1}{2}}-1\,,
\end{equation}
where $M$ and $R$ represent the mass and radius of the NS, respectively. For example, assuming a NS with a mass of $1.4\,M_{\odot}$ and a radius of $10\,\rm km$, the gravitational redshift factor is 0.306.

To estimate the redshift factor caused by the binary system, we neglect the spin of the companion and crudely treat the NS as a test particle. We assume that the observer is located in the orbital plane and at an orbital phase of $\phi=0$ when the NS is closest to the observer. We select several special points to discuss the frequency shift. At $\phi=0, \pi$, the redshift factor is given by:
\begin{equation}
z_{\rm binary}=\left(1-\frac{3GM_c}{Dc^{2}}\right)^{-1/2}-1\,,
\end{equation}
where $M_c$ represents the total mass of the binary system, and $D$ represents the distance between the observer and the binary system. This equation takes into account both the gravitational redshift and the transverse Doppler effect caused by the motion of the NS in the binary system.

At $\phi=\pm \pi/2$, the redshift factor for the binary system is given by
\begin{equation}
z_{\rm binary}=\left(1-\frac{3GM_c}{Dc^{2}}\right)^{-1/2}\left( 1\pm \left(\frac{Dc^2}{GM}-2\right)^{-1/2}\right)-1\,.
\end{equation}
For a binary system with $M_c=10\,M_{\odot}$, $D=100\,\rm km$, $M=1.4\,M_{\odot}$, and $R=10\,\rm km$, the redshift factor $z_{\rm binary}(\phi=0)=z_{\rm binary}(\phi=\pi/2)=0.34$, $z_{\rm binary}(\phi=\pi/2)=0.95$, and $z_{\rm binary}(\phi=-\pi/2)=-0.27$. The redshift $z_{\rm binary}$ is largest at $\phi=\pi/2$. The frequency is blueshifted at $\phi=-\pi/2$ since the Doppler effect dominates over the gravitational redshift.

The relation between the observed frequency $f_{\rm obs}$ and the mode frequency in the local frame of the NS, $f$, can be approximated by:
\begin{equation}
f_{\rm obs}=(1+z_{\rm ns})^{-1}(1+z_{\rm binary})^{-1}f\,.
\end{equation}
The largest redshift corresponds to $f_{\rm obs}(\phi=\pi/2)=0.393f$.

\section{Conclusions and Discussions}
\label{sec:dc}
In this paper, we introduced the scenario of a tidal induced starquake for a solid strange star prior to the merger with a compact companion. As the binary inspirals closer, the tidal field will gradually deform the solid strange star, resulting in the accumulation of the elastic energy. A giant starquake might be triggered when the binary is close enough and the stress exceeds a critical value. This scenario is quite similar to the starquake model of pulsar glitches, in which the elastic energy is accumulated as the star deforms due to the spinning down process. 

In particular, we demonstrated that such giant quake before merger could release sufficient energy to explain the precursor observation of GRB211211A. In the sequential quakes model, which we consider only part of the star suffers the starquake, the released energy is still enough at a separation larger than the ISCO radius.
Moreover, the torsional mode frequencies are estimated by regarding the entire solid strange star as a homogeneous, isotropic elastic spheroid with uniform density and shear modulus. The result depends on the final shear modulus when starquake happens (i.e., after the failure of the solid structure takes place) and could be consistent with the observation with a reasonable range of the final shear modulus, without contradicting the energy budget calculations.
Our result favors a BH-NS scenario compared with BNS case, due to much larger parameter space for the consideration of the energy budget. The BH-NS scenario is also more consistent with the event rates. sGRBs with confirmed precursor observation are quite rare, just as BH-NS binary systems are much fewer compared with BNS systems. We will esitmate the event rates in the future and try to verify this argument.

We are expecting to test the model presented in this paper in the future, especially by combining the observations of gravitational and electro-magnetic signals.
The LVK-O4 observing run (e.g., \citep{2023arXiv230204147C}) would start, and China's mega-facilities, especially the GECAM~\citep{2020SSPMA..50l9508L}, the HXMT~\citep{2018SCPMA..61c1011L}, as well as the planned EP~\citep{2022hxga.book...86Y}, would work. 
The model we proposed might then be soon falsified, but, conversely, would show its particular reasoning style in the coming years.

\acknowledgements
We thank the Computational Relativistic Astrophysics division in Albert Einstein Institute (Potsdam) for helpful discussions on the kilonova scenario of BH-NS mergers. 
This work was supported by the National SKA Program
of China (2020SKA01201000, 2020SKA0120300). E. Zhou is supported by NSFC Grant No. 12203017. Y. Gao and L. Shao are supported by the National Natural Science Foundation of China (11975027, 11991053) and the Max Planck Partner Group Program funded by the Max Planck Society. S. Xiong and Z. Zhang are supported by the national Key R\&D Program of China (2021YFA0718500),  National Natural Science Foundation of China (Grant No. 12273042). X. Lai accknowledges the support by the Young Top-notch Talent Cultivation Program of Hubei Province. H. Yue is supported by NSFC Grant No. 42174059 and S. Yi acknowledges the support from the Chinese Academy of Sciences (Grant No. E329A3M1).

\bibliographystyle{apsrev4-1}
\bibliography{aeireferences}

\begin{thebibliography}{56}%
\makeatletter
\providecommand \@ifxundefined [1]{%
 \@ifx{#1\undefined}
}%
\providecommand \@ifnum [1]{%
 \ifnum #1\expandafter \@firstoftwo
 \else \expandafter \@secondoftwo
 \fi
}%
\providecommand \@ifx [1]{%
 \ifx #1\expandafter \@firstoftwo
 \else \expandafter \@secondoftwo
 \fi
}%
\providecommand \natexlab [1]{#1}%
\providecommand \enquote  [1]{``#1''}%
\providecommand \bibnamefont  [1]{#1}%
\providecommand \bibfnamefont [1]{#1}%
\providecommand \citenamefont [1]{#1}%
\providecommand \href@noop [0]{\@secondoftwo}%
\providecommand \href [0]{\begingroup \@sanitize@url \@href}%
\providecommand \@href[1]{\@@startlink{#1}\@@href}%
\providecommand \@@href[1]{\endgroup#1\@@endlink}%
\providecommand \@sanitize@url [0]{\catcode `\\12\catcode `\$12\catcode
  `\&12\catcode `\#12\catcode `\^12\catcode `\_12\catcode `\%12\relax}%
\providecommand \@@startlink[1]{}%
\providecommand \@@endlink[0]{}%
\providecommand \url  [0]{\begingroup\@sanitize@url \@url }%
\providecommand \@url [1]{\endgroup\@href {#1}{\urlprefix }}%
\providecommand \urlprefix  [0]{URL }%
\providecommand \Eprint [0]{\href }%
\providecommand \doibase [0]{http://dx.doi.org/}%
\providecommand \selectlanguage [0]{\@gobble}%
\providecommand \bibinfo  [0]{\@secondoftwo}%
\providecommand \bibfield  [0]{\@secondoftwo}%
\providecommand \translation [1]{[#1]}%
\providecommand \BibitemOpen [0]{}%
\providecommand \bibitemStop [0]{}%
\providecommand \bibitemNoStop [0]{.\EOS\space}%
\providecommand \EOS [0]{\spacefactor3000\relax}%
\providecommand \BibitemShut  [1]{\csname bibitem#1\endcsname}%
\let\auto@bib@innerbib\@empty
\bibitem [{\citenamefont {{Epelbaum}}\ \emph {et~al.}(2009)\citenamefont
  {{Epelbaum}}, \citenamefont {{Hammer}},\ and\ \citenamefont
  {{Mei{\ss}ner}}}]{2009RvMP...81.1773E}%
  \BibitemOpen
  \bibfield  {author} {\bibinfo {author} {\bibfnamefont {E.}~\bibnamefont
  {{Epelbaum}}}, \bibinfo {author} {\bibfnamefont {H.~W.}\ \bibnamefont
  {{Hammer}}}, \ and\ \bibinfo {author} {\bibfnamefont {U.-G.}\ \bibnamefont
  {{Mei{\ss}ner}}},\ }\href {\doibase 10.1103/RevModPhys.81.1773} {\bibfield
  {journal} {\bibinfo  {journal} {Reviews of Modern Physics}\ }\textbf
  {\bibinfo {volume} {81}},\ \bibinfo {pages} {1773} (\bibinfo {year}
  {2009})},\ \Eprint {http://arxiv.org/abs/0811.1338} {arXiv:0811.1338
  [nucl-th]} \BibitemShut {NoStop}%
\bibitem [{\citenamefont {{Baiotti}}(2019)}]{2019PrPNP.10903714B}%
  \BibitemOpen
  \bibfield  {author} {\bibinfo {author} {\bibfnamefont {L.}~\bibnamefont
  {{Baiotti}}},\ }\href {\doibase 10.1016/j.ppnp.2019.103714} {\bibfield
  {journal} {\bibinfo  {journal} {Progress in Particle and Nuclear Physics}\
  }\textbf {\bibinfo {volume} {109}},\ \bibinfo {eid} {103714} (\bibinfo {year}
  {2019})},\ \Eprint {http://arxiv.org/abs/1907.08534} {arXiv:1907.08534
  [astro-ph.HE]} \BibitemShut {NoStop}%
\bibitem [{\citenamefont {{Xu}}(2003)}]{Xu2003}%
  \BibitemOpen
  \bibfield  {author} {\bibinfo {author} {\bibfnamefont {R.~X.}\ \bibnamefont
  {{Xu}}},\ }\href {\doibase 10.1086/379209} {\bibfield  {journal} {\bibinfo
  {journal} {Astrophys. J. Lett.}\ }\textbf {\bibinfo {volume} {596}},\
  \bibinfo {pages} {L59} (\bibinfo {year} {2003})},\ \Eprint
  {http://arxiv.org/abs/astro-ph/0302165} {astro-ph/0302165} \BibitemShut
  {NoStop}%
\bibitem [{\citenamefont {{Xu}}\ \emph {et~al.}(2006)\citenamefont {{Xu}},
  \citenamefont {{Tao}},\ and\ \citenamefont {{Yang}}}]{2006MNRAS.373L..85X}%
  \BibitemOpen
  \bibfield  {author} {\bibinfo {author} {\bibfnamefont {R.~X.}\ \bibnamefont
  {{Xu}}}, \bibinfo {author} {\bibfnamefont {D.~J.}\ \bibnamefont {{Tao}}}, \
  and\ \bibinfo {author} {\bibfnamefont {Y.}~\bibnamefont {{Yang}}},\ }\href
  {\doibase 10.1111/j.1745-3933.2006.00248.x} {\bibfield  {journal} {\bibinfo
  {journal} {\mnras}\ }\textbf {\bibinfo {volume} {373}},\ \bibinfo {pages}
  {L85} (\bibinfo {year} {2006})},\ \Eprint
  {http://arxiv.org/abs/astro-ph/0607106} {arXiv:astro-ph/0607106 [astro-ph]}
  \BibitemShut {NoStop}%
\bibitem [{\citenamefont {{Lai}}\ \emph {et~al.}(2019)\citenamefont {{Lai}},
  \citenamefont {{Zhou}},\ and\ \citenamefont {{Xu}}}]{2019EPJA...55...60L}%
  \BibitemOpen
  \bibfield  {author} {\bibinfo {author} {\bibfnamefont {X.}~\bibnamefont
  {{Lai}}}, \bibinfo {author} {\bibfnamefont {E.}~\bibnamefont {{Zhou}}}, \
  and\ \bibinfo {author} {\bibfnamefont {R.}~\bibnamefont {{Xu}}},\ }\href
  {\doibase 10.1140/epja/i2019-12720-8} {\bibfield  {journal} {\bibinfo
  {journal} {European Physical Journal A}\ }\textbf {\bibinfo {volume} {55}},\
  \bibinfo {eid} {60} (\bibinfo {year} {2019})},\ \Eprint
  {http://arxiv.org/abs/1811.00193} {arXiv:1811.00193 [astro-ph.HE]}
  \BibitemShut {NoStop}%
\bibitem [{\citenamefont {{Coupechoux}}\ \emph {et~al.}(2023)\citenamefont
  {{Coupechoux}}, \citenamefont {{Chierici}}, \citenamefont {{Hansen}},
  \citenamefont {{Margueron}}, \citenamefont {{Somasundaram}},\ and\
  \citenamefont {{Sordini}}}]{2023arXiv230204147C}%
  \BibitemOpen
  \bibfield  {author} {\bibinfo {author} {\bibfnamefont {J.~F.}\ \bibnamefont
  {{Coupechoux}}}, \bibinfo {author} {\bibfnamefont {R.}~\bibnamefont
  {{Chierici}}}, \bibinfo {author} {\bibfnamefont {H.}~\bibnamefont
  {{Hansen}}}, \bibinfo {author} {\bibfnamefont {J.}~\bibnamefont
  {{Margueron}}}, \bibinfo {author} {\bibfnamefont {R.}~\bibnamefont
  {{Somasundaram}}}, \ and\ \bibinfo {author} {\bibfnamefont {V.}~\bibnamefont
  {{Sordini}}},\ }\href {\doibase 10.48550/arXiv.2302.04147} {\bibfield
  {journal} {\bibinfo  {journal} {arXiv e-prints}\ ,\ \bibinfo {eid}
  {arXiv:2302.04147}} (\bibinfo {year} {2023})},\ \Eprint
  {http://arxiv.org/abs/2302.04147} {arXiv:2302.04147 [astro-ph.HE]}
  \BibitemShut {NoStop}%
\bibitem [{\citenamefont {{The LIGO Scientific Collaboration}}\ and\
  \citenamefont {{The Virgo Collaboration}}(2017)}]{Abbott2017}%
  \BibitemOpen
  \bibfield  {author} {\bibinfo {author} {\bibnamefont {{The LIGO Scientific
  Collaboration}}}\ and\ \bibinfo {author} {\bibnamefont {{The Virgo
  Collaboration}}} (\bibinfo {collaboration} {LIGO Scientific Collaboration and
  Virgo Collaboration}),\ }\href {\doibase 10.1103/PhysRevLett.119.161101}
  {\bibfield  {journal} {\bibinfo  {journal} {Phys. Rev. Lett.}\ }\textbf
  {\bibinfo {volume} {119}},\ \bibinfo {pages} {161101} (\bibinfo {year}
  {2017})}\BibitemShut {NoStop}%
\bibitem [{\citenamefont {{The LIGO Scientific Collaboration}}\ \emph
  {et~al.}(2017)\citenamefont {{The LIGO Scientific Collaboration}},
  \citenamefont {{the Virgo Collaboration}}, \citenamefont {{Abbott}},
  \citenamefont {{Abbott}}, \citenamefont {{Abbott}}, \citenamefont
  {{Acernese}}, \citenamefont {{Ackley}}, \citenamefont {a{Adams}},
  \citenamefont {{Adams}}, \citenamefont {{Addesso}},\ and\ \citenamefont
  {et~al.}}]{Abbott2017b}%
  \BibitemOpen
  \bibfield  {author} {\bibinfo {author} {\bibnamefont {{The LIGO Scientific
  Collaboration}}}, \bibinfo {author} {\bibnamefont {{the Virgo
  Collaboration}}}, \bibinfo {author} {\bibfnamefont {B.~P.}\ \bibnamefont
  {{Abbott}}}, \bibinfo {author} {\bibfnamefont {R.}~\bibnamefont {{Abbott}}},
  \bibinfo {author} {\bibfnamefont {T.~D.}\ \bibnamefont {{Abbott}}}, \bibinfo
  {author} {\bibfnamefont {F.}~\bibnamefont {{Acernese}}}, \bibinfo {author}
  {\bibfnamefont {K.}~\bibnamefont {{Ackley}}}, \bibinfo {author}
  {\bibfnamefont {C.}~\bibnamefont {a{Adams}}}, \bibinfo {author}
  {\bibfnamefont {T.}~\bibnamefont {{Adams}}}, \bibinfo {author} {\bibfnamefont
  {P.}~\bibnamefont {{Addesso}}}, \ and\ \bibinfo {author} {\bibnamefont
  {et~al.}} (\bibinfo {collaboration} {LIGO Scientific Collaboration and Virgo
  Collaboration}),\ }\href {http://stacks.iop.org/2041-8205/848/i=2/a=L12}
  {\bibfield  {journal} {\bibinfo  {journal} {Astrophys. J. Lett.}\ }\textbf
  {\bibinfo {volume} {848}},\ \bibinfo {pages} {L12} (\bibinfo {year}
  {2017})}\BibitemShut {NoStop}%
\bibitem [{\citenamefont {{LIGO Scientific Collaboration}}\ \emph
  {et~al.}(2017)\citenamefont {{LIGO Scientific Collaboration}}, \citenamefont
  {{Virgo Collaboration}}, \citenamefont {{Gamma-Ray Burst Monitor}},\ and\
  \citenamefont {{INTEGRAL}}}]{Abbott2017d}%
  \BibitemOpen
  \bibfield  {author} {\bibinfo {author} {\bibnamefont {{LIGO Scientific
  Collaboration}}}, \bibinfo {author} {\bibnamefont {{Virgo Collaboration}}},
  \bibinfo {author} {\bibfnamefont {F.}~\bibnamefont {{Gamma-Ray Burst
  Monitor}}}, \ and\ \bibinfo {author} {\bibnamefont {{INTEGRAL}}},\ }\href
  {http://stacks.iop.org/2041-8205/848/i=2/a=L13} {\bibfield  {journal}
  {\bibinfo  {journal} {Astrophys. J. Lett.}\ }\textbf {\bibinfo {volume}
  {848}},\ \bibinfo {pages} {L13} (\bibinfo {year} {2017})},\ \Eprint
  {http://arxiv.org/abs/1710.05834} {arXiv:1710.05834 [astro-ph.HE]}
  \BibitemShut {NoStop}%
\bibitem [{\citenamefont {{Narayan}}\ \emph {et~al.}(1992)\citenamefont
  {{Narayan}}, \citenamefont {{Paczynski}},\ and\ \citenamefont
  {{Piran}}}]{Narayan92}%
  \BibitemOpen
  \bibfield  {author} {\bibinfo {author} {\bibfnamefont {R.}~\bibnamefont
  {{Narayan}}}, \bibinfo {author} {\bibfnamefont {B.}~\bibnamefont
  {{Paczynski}}}, \ and\ \bibinfo {author} {\bibfnamefont {T.}~\bibnamefont
  {{Piran}}},\ }\href {\doibase 10.1086/186493} {\bibfield  {journal} {\bibinfo
   {journal} {Astrophys. J. Lett.}\ }\textbf {\bibinfo {volume} {395}},\
  \bibinfo {pages} {L83} (\bibinfo {year} {1992})},\ \Eprint
  {http://arxiv.org/abs/astro-ph/9204001} {astro-ph/9204001} \BibitemShut
  {NoStop}%
\bibitem [{\citenamefont {{Ruiz}}\ \emph {et~al.}(2018)\citenamefont {{Ruiz}},
  \citenamefont {{Shapiro}},\ and\ \citenamefont {{Tsokaros}}}]{Ruiz2017}%
  \BibitemOpen
  \bibfield  {author} {\bibinfo {author} {\bibfnamefont {M.}~\bibnamefont
  {{Ruiz}}}, \bibinfo {author} {\bibfnamefont {S.~L.}\ \bibnamefont
  {{Shapiro}}}, \ and\ \bibinfo {author} {\bibfnamefont {A.}~\bibnamefont
  {{Tsokaros}}},\ }\href {\doibase 10.1103/PhysRevD.97.021501} {\bibfield
  {journal} {\bibinfo  {journal} {Phys. Rev. D}\ }\textbf {\bibinfo {volume}
  {97}},\ \bibinfo {eid} {021501} (\bibinfo {year} {2018})},\ \Eprint
  {http://arxiv.org/abs/1711.00473} {arXiv:1711.00473 [astro-ph.HE]}
  \BibitemShut {NoStop}%
\bibitem [{\citenamefont {{Rezzolla}}\ \emph {et~al.}(2018)\citenamefont
  {{Rezzolla}}, \citenamefont {{Most}},\ and\ \citenamefont
  {{Weih}}}]{Rezzolla2017}%
  \BibitemOpen
  \bibfield  {author} {\bibinfo {author} {\bibfnamefont {L.}~\bibnamefont
  {{Rezzolla}}}, \bibinfo {author} {\bibfnamefont {E.~R.}\ \bibnamefont
  {{Most}}}, \ and\ \bibinfo {author} {\bibfnamefont {L.~R.}\ \bibnamefont
  {{Weih}}},\ }\href {\doibase 10.3847/2041-8213/aaa401} {\bibfield  {journal}
  {\bibinfo  {journal} {Astrophys. J. Lett.}\ }\textbf {\bibinfo {volume}
  {852}},\ \bibinfo {eid} {L25} (\bibinfo {year} {2018})},\ \Eprint
  {http://arxiv.org/abs/1711.00314} {arXiv:1711.00314 [astro-ph.HE]}
  \BibitemShut {NoStop}%
\bibitem [{\citenamefont {{Bauswein}}\ \emph {et~al.}(2017)\citenamefont
  {{Bauswein}}, \citenamefont {{Just}}, \citenamefont {{Janka}},\ and\
  \citenamefont {{Stergioulas}}}]{Bauswein2017b}%
  \BibitemOpen
  \bibfield  {author} {\bibinfo {author} {\bibfnamefont {A.}~\bibnamefont
  {{Bauswein}}}, \bibinfo {author} {\bibfnamefont {O.}~\bibnamefont {{Just}}},
  \bibinfo {author} {\bibfnamefont {H.-T.}\ \bibnamefont {{Janka}}}, \ and\
  \bibinfo {author} {\bibfnamefont {N.}~\bibnamefont {{Stergioulas}}},\ }\href
  {\doibase 10.3847/2041-8213/aa9994} {\bibfield  {journal} {\bibinfo
  {journal} {Astrophys. J. Lett.}\ }\textbf {\bibinfo {volume} {850}},\
  \bibinfo {eid} {L34} (\bibinfo {year} {2017})},\ \Eprint
  {http://arxiv.org/abs/1710.06843} {arXiv:1710.06843 [astro-ph.HE]}
  \BibitemShut {NoStop}%
\bibitem [{\citenamefont {{Margalit}}\ and\ \citenamefont
  {{Metzger}}(2017)}]{Margalit2017}%
  \BibitemOpen
  \bibfield  {author} {\bibinfo {author} {\bibfnamefont {B.}~\bibnamefont
  {{Margalit}}}\ and\ \bibinfo {author} {\bibfnamefont {B.~D.}\ \bibnamefont
  {{Metzger}}},\ }\href {\doibase 10.3847/2041-8213/aa991c} {\bibfield
  {journal} {\bibinfo  {journal} {Astrophys. J. Lett.}\ }\textbf {\bibinfo
  {volume} {850}},\ \bibinfo {eid} {L19} (\bibinfo {year} {2017})},\ \Eprint
  {http://arxiv.org/abs/1710.05938} {arXiv:1710.05938 [astro-ph.HE]}
  \BibitemShut {NoStop}%
\bibitem [{\citenamefont {{Shibata}}\ \emph {et~al.}(2019)\citenamefont
  {{Shibata}}, \citenamefont {{Zhou}}, \citenamefont {{Kiuchi}},\ and\
  \citenamefont {{Fujibayashi}}}]{Shibata2019}%
  \BibitemOpen
  \bibfield  {author} {\bibinfo {author} {\bibfnamefont {M.}~\bibnamefont
  {{Shibata}}}, \bibinfo {author} {\bibfnamefont {E.}~\bibnamefont {{Zhou}}},
  \bibinfo {author} {\bibfnamefont {K.}~\bibnamefont {{Kiuchi}}}, \ and\
  \bibinfo {author} {\bibfnamefont {S.}~\bibnamefont {{Fujibayashi}}},\ }\href
  {\doibase 10.1103/PhysRevD.100.023015} {\bibfield  {journal} {\bibinfo
  {journal} {\prd}\ }\textbf {\bibinfo {volume} {100}},\ \bibinfo {eid}
  {023015} (\bibinfo {year} {2019})},\ \Eprint
  {http://arxiv.org/abs/1905.03656} {arXiv:1905.03656 [astro-ph.HE]}
  \BibitemShut {NoStop}%
\bibitem [{\citenamefont {{Kiuchi}}\ \emph {et~al.}(2019)\citenamefont
  {{Kiuchi}}, \citenamefont {{Kyutoku}}, \citenamefont {{Shibata}},\ and\
  \citenamefont {{Taniguchi}}}]{Kiuchi2019}%
  \BibitemOpen
  \bibfield  {author} {\bibinfo {author} {\bibfnamefont {K.}~\bibnamefont
  {{Kiuchi}}}, \bibinfo {author} {\bibfnamefont {K.}~\bibnamefont {{Kyutoku}}},
  \bibinfo {author} {\bibfnamefont {M.}~\bibnamefont {{Shibata}}}, \ and\
  \bibinfo {author} {\bibfnamefont {K.}~\bibnamefont {{Taniguchi}}},\ }\href
  {\doibase 10.3847/2041-8213/ab1e45} {\bibfield  {journal} {\bibinfo
  {journal} {\apjl}\ }\textbf {\bibinfo {volume} {876}},\ \bibinfo {eid} {L31}
  (\bibinfo {year} {2019})},\ \Eprint {http://arxiv.org/abs/1903.01466}
  {arXiv:1903.01466 [astro-ph.HE]} \BibitemShut {NoStop}%
\bibitem [{\citenamefont {{Abbott}}\ \emph {et~al.}(2017)\citenamefont
  {{Abbott}}, \citenamefont {{Abbott}}, \citenamefont {{Abbott}}, \citenamefont
  {{Acernese}}, \citenamefont {{Ackley}}, \citenamefont {{Adams}},
  \citenamefont {{Adams}}, \citenamefont {{Addesso}}, \citenamefont
  {{Adhikari}}, \citenamefont {{Adya}},\ and\ \citenamefont
  {et~al.}}]{Abbott2017c}%
  \BibitemOpen
  \bibfield  {author} {\bibinfo {author} {\bibfnamefont {B.~P.}\ \bibnamefont
  {{Abbott}}}, \bibinfo {author} {\bibfnamefont {R.}~\bibnamefont {{Abbott}}},
  \bibinfo {author} {\bibfnamefont {T.~D.}\ \bibnamefont {{Abbott}}}, \bibinfo
  {author} {\bibfnamefont {F.}~\bibnamefont {{Acernese}}}, \bibinfo {author}
  {\bibfnamefont {K.}~\bibnamefont {{Ackley}}}, \bibinfo {author}
  {\bibfnamefont {C.}~\bibnamefont {{Adams}}}, \bibinfo {author} {\bibfnamefont
  {T.}~\bibnamefont {{Adams}}}, \bibinfo {author} {\bibfnamefont
  {P.}~\bibnamefont {{Addesso}}}, \bibinfo {author} {\bibfnamefont {R.~X.}\
  \bibnamefont {{Adhikari}}}, \bibinfo {author} {\bibfnamefont {V.~B.}\
  \bibnamefont {{Adya}}}, \ and\ \bibinfo {author} {\bibnamefont {et~al.}},\
  }\href {\doibase 10.3847/2041-8213/aa9478} {\bibfield  {journal} {\bibinfo
  {journal} {Astrophys. J. Lett.}\ }\textbf {\bibinfo {volume} {850}},\
  \bibinfo {eid} {L39} (\bibinfo {year} {2017})},\ \Eprint
  {http://arxiv.org/abs/1710.05836} {arXiv:1710.05836 [astro-ph.HE]}
  \BibitemShut {NoStop}%
\bibitem [{\citenamefont {{Eichler}}\ \emph {et~al.}(1989)\citenamefont
  {{Eichler}}, \citenamefont {{Livio}}, \citenamefont {{Piran}},\ and\
  \citenamefont {{Schramm}}}]{Eichler89}%
  \BibitemOpen
  \bibfield  {author} {\bibinfo {author} {\bibfnamefont {D.}~\bibnamefont
  {{Eichler}}}, \bibinfo {author} {\bibfnamefont {M.}~\bibnamefont {{Livio}}},
  \bibinfo {author} {\bibfnamefont {T.}~\bibnamefont {{Piran}}}, \ and\
  \bibinfo {author} {\bibfnamefont {D.~N.}\ \bibnamefont {{Schramm}}},\ }\href
  {\doibase 10.1038/340126a0} {\bibfield  {journal} {\bibinfo  {journal}
  {Nature}\ }\textbf {\bibinfo {volume} {340}},\ \bibinfo {pages} {126}
  (\bibinfo {year} {1989})}\BibitemShut {NoStop}%
\bibitem [{\citenamefont {{Mangan}}\ \emph {et~al.}(2021)\citenamefont
  {{Mangan}}, \citenamefont {{Dunwoody}}, \citenamefont {{Meegan}},\ and\
  \citenamefont {{Fermi GBM Team}}}]{Fermi2021}%
  \BibitemOpen
  \bibfield  {author} {\bibinfo {author} {\bibfnamefont {J.}~\bibnamefont
  {{Mangan}}}, \bibinfo {author} {\bibfnamefont {R.}~\bibnamefont
  {{Dunwoody}}}, \bibinfo {author} {\bibfnamefont {C.}~\bibnamefont
  {{Meegan}}}, \ and\ \bibinfo {author} {\bibnamefont {{Fermi GBM Team}}},\
  }\href@noop {} {\bibfield  {journal} {\bibinfo  {journal} {GRB Coordinates
  Network}\ }\textbf {\bibinfo {volume} {31210}},\ \bibinfo {pages} {1}
  (\bibinfo {year} {2021})}\BibitemShut {NoStop}%
\bibitem [{\citenamefont {{D'Ai}}\ \emph {et~al.}(2021)\citenamefont {{D'Ai}},
  \citenamefont {{Ambrosi}}, \citenamefont {{D'Elia}}, \citenamefont {{Gropp}},
  \citenamefont {{Kennea}}, \citenamefont {{Kuin}}, \citenamefont {{Lien}},
  \citenamefont {{Marshall}}, \citenamefont {{Page}}, \citenamefont {{Palmer}},
  \citenamefont {{Parsotan}}, \citenamefont {{Sbarufatti}},\ and\ \citenamefont
  {{Neil Gehrels Swift Observatory Team}}}]{swift2021}%
  \BibitemOpen
  \bibfield  {author} {\bibinfo {author} {\bibfnamefont {A.}~\bibnamefont
  {{D'Ai}}}, \bibinfo {author} {\bibfnamefont {E.}~\bibnamefont {{Ambrosi}}},
  \bibinfo {author} {\bibfnamefont {V.}~\bibnamefont {{D'Elia}}}, \bibinfo
  {author} {\bibfnamefont {J.~D.}\ \bibnamefont {{Gropp}}}, \bibinfo {author}
  {\bibfnamefont {J.~A.}\ \bibnamefont {{Kennea}}}, \bibinfo {author}
  {\bibfnamefont {N.~P.~M.}\ \bibnamefont {{Kuin}}}, \bibinfo {author}
  {\bibfnamefont {A.~Y.}\ \bibnamefont {{Lien}}}, \bibinfo {author}
  {\bibfnamefont {F.~E.}\ \bibnamefont {{Marshall}}}, \bibinfo {author}
  {\bibfnamefont {K.~L.}\ \bibnamefont {{Page}}}, \bibinfo {author}
  {\bibfnamefont {D.~M.}\ \bibnamefont {{Palmer}}}, \bibinfo {author}
  {\bibfnamefont {T.~M.}\ \bibnamefont {{Parsotan}}}, \bibinfo {author}
  {\bibfnamefont {B.}~\bibnamefont {{Sbarufatti}}}, \ and\ \bibinfo {author}
  {\bibnamefont {{Neil Gehrels Swift Observatory Team}}},\ }\href@noop {}
  {\bibfield  {journal} {\bibinfo  {journal} {GRB Coordinates Network}\
  }\textbf {\bibinfo {volume} {31202}},\ \bibinfo {pages} {1} (\bibinfo {year}
  {2021})}\BibitemShut {NoStop}%
\bibitem [{\citenamefont {{Zhang}}\ \emph {et~al.}(2021)\citenamefont
  {{Zhang}}, \citenamefont {{Xiong}}, \citenamefont {{Li}}, \citenamefont
  {{Cai}}, \citenamefont {{Luo}}, \citenamefont {{Xiao}}, \citenamefont
  {{Liu}}, \citenamefont {{Xue}}, \citenamefont {{Yi}}, \citenamefont
  {{Zheng}}, \citenamefont {{Li}}, \citenamefont {{Li}}, \citenamefont
  {{Liao}}, \citenamefont {{Song}}, \citenamefont {{Xiong}}, \citenamefont
  {{Liu}}, \citenamefont {{Li}}, \citenamefont {{Li}}, \citenamefont {{Chang}},
  \citenamefont {{Zhang}}, \citenamefont {{Zhang}}, \citenamefont {{Lu}},
  \citenamefont {{Zou}}, \citenamefont {{Jin}}, \citenamefont {{Zhang}},
  \citenamefont {{Li}}, \citenamefont {{Lu}}, \citenamefont {{Song}},
  \citenamefont {{Wu}}, \citenamefont {{Xu}}, \citenamefont {{Zhang}},\ and\
  \citenamefont {{Insight-HXMT Team}}}]{insight2021}%
  \BibitemOpen
  \bibfield  {author} {\bibinfo {author} {\bibfnamefont {Y.~Q.}\ \bibnamefont
  {{Zhang}}}, \bibinfo {author} {\bibfnamefont {S.~L.}\ \bibnamefont
  {{Xiong}}}, \bibinfo {author} {\bibfnamefont {X.~B.}\ \bibnamefont {{Li}}},
  \bibinfo {author} {\bibfnamefont {C.}~\bibnamefont {{Cai}}}, \bibinfo
  {author} {\bibfnamefont {Q.}~\bibnamefont {{Luo}}}, \bibinfo {author}
  {\bibfnamefont {S.}~\bibnamefont {{Xiao}}}, \bibinfo {author} {\bibfnamefont
  {J.~C.}\ \bibnamefont {{Liu}}}, \bibinfo {author} {\bibfnamefont {W.~C.}\
  \bibnamefont {{Xue}}}, \bibinfo {author} {\bibfnamefont {Q.~B.}\ \bibnamefont
  {{Yi}}}, \bibinfo {author} {\bibfnamefont {C.}~\bibnamefont {{Zheng}},
  \bibfnamefont {Y.~Huang}}, \bibinfo {author} {\bibfnamefont {C.~K.}\
  \bibnamefont {{Li}}}, \bibinfo {author} {\bibfnamefont {G.}~\bibnamefont
  {{Li}}}, \bibinfo {author} {\bibfnamefont {J.~Y.}\ \bibnamefont {{Liao}}},
  \bibinfo {author} {\bibfnamefont {X.~Y.}\ \bibnamefont {{Song}}}, \bibinfo
  {author} {\bibfnamefont {S.~L.}\ \bibnamefont {{Xiong}}}, \bibinfo {author}
  {\bibfnamefont {C.~Z.}\ \bibnamefont {{Liu}}}, \bibinfo {author}
  {\bibfnamefont {X.~F.}\ \bibnamefont {{Li}}}, \bibinfo {author}
  {\bibfnamefont {Z.~W.}\ \bibnamefont {{Li}}}, \bibinfo {author}
  {\bibfnamefont {Z.}~\bibnamefont {{Chang}}}, \bibinfo {author} {\bibfnamefont
  {A.~M.}\ \bibnamefont {{Zhang}}}, \bibinfo {author} {\bibfnamefont {Y.~F.}\
  \bibnamefont {{Zhang}}}, \bibinfo {author} {\bibfnamefont {X.~F.}\
  \bibnamefont {{Lu}}}, \bibinfo {author} {\bibfnamefont {C.~L.}\ \bibnamefont
  {{Zou}}}, \bibinfo {author} {\bibfnamefont {Y.~J.}\ \bibnamefont {{Jin}}},
  \bibinfo {author} {\bibfnamefont {Z.}~\bibnamefont {{Zhang}}}, \bibinfo
  {author} {\bibfnamefont {T.~P.}\ \bibnamefont {{Li}}}, \bibinfo {author}
  {\bibfnamefont {F.~J.}\ \bibnamefont {{Lu}}}, \bibinfo {author}
  {\bibfnamefont {L.~M.}\ \bibnamefont {{Song}}}, \bibinfo {author}
  {\bibfnamefont {M.}~\bibnamefont {{Wu}}}, \bibinfo {author} {\bibfnamefont
  {Y.~P.}\ \bibnamefont {{Xu}}}, \bibinfo {author} {\bibfnamefont {S.~N.}\
  \bibnamefont {{Zhang}}}, \ and\ \bibinfo {author} {\bibnamefont
  {{Insight-HXMT Team}}},\ }\href@noop {} {\bibfield  {journal} {\bibinfo
  {journal} {GRB Coordinates Network}\ }\textbf {\bibinfo {volume} {31236}},\
  \bibinfo {pages} {1} (\bibinfo {year} {2021})}\BibitemShut {NoStop}%
\bibitem [{\citenamefont {{Rastinejad}}\ \emph {et~al.}(2022)\citenamefont
  {{Rastinejad}}, \citenamefont {{Gompertz}}, \citenamefont {{Levan}},
  \citenamefont {{Fong}}, \citenamefont {{Nicholl}}, \citenamefont {{Lamb}},
  \citenamefont {{Malesani}}, \citenamefont {{Nugent}}, \citenamefont
  {{Oates}}, \citenamefont {{Tanvir}}, \citenamefont {{de Ugarte Postigo}},
  \citenamefont {{Kilpatrick}}, \citenamefont {{Moore}}, \citenamefont
  {{Metzger}}, \citenamefont {{Ravasio}}, \citenamefont {{Rossi}},
  \citenamefont {{Schroeder}}, \citenamefont {{Jencson}}, \citenamefont
  {{Sand}}, \citenamefont {{Smith}}, \citenamefont {{Ag{\"u}{\'\i}
  Fern{\'a}ndez}}, \citenamefont {{Berger}}, \citenamefont {{Blanchard}},
  \citenamefont {{Chornock}}, \citenamefont {{Cobb}}, \citenamefont {{De
  Pasquale}}, \citenamefont {{Fynbo}}, \citenamefont {{Izzo}}, \citenamefont
  {{Kann}}, \citenamefont {{Laskar}}, \citenamefont {{Marini}}, \citenamefont
  {{Paterson}}, \citenamefont {{Escorial}}, \citenamefont {{Sears}},\ and\
  \citenamefont {{Th{\"o}ne}}}]{Rastinejad2022}%
  \BibitemOpen
  \bibfield  {author} {\bibinfo {author} {\bibfnamefont {J.~C.}\ \bibnamefont
  {{Rastinejad}}}, \bibinfo {author} {\bibfnamefont {B.~P.}\ \bibnamefont
  {{Gompertz}}}, \bibinfo {author} {\bibfnamefont {A.~J.}\ \bibnamefont
  {{Levan}}}, \bibinfo {author} {\bibfnamefont {W.-f.}\ \bibnamefont {{Fong}}},
  \bibinfo {author} {\bibfnamefont {M.}~\bibnamefont {{Nicholl}}}, \bibinfo
  {author} {\bibfnamefont {G.~P.}\ \bibnamefont {{Lamb}}}, \bibinfo {author}
  {\bibfnamefont {D.~B.}\ \bibnamefont {{Malesani}}}, \bibinfo {author}
  {\bibfnamefont {A.~E.}\ \bibnamefont {{Nugent}}}, \bibinfo {author}
  {\bibfnamefont {S.~R.}\ \bibnamefont {{Oates}}}, \bibinfo {author}
  {\bibfnamefont {N.~R.}\ \bibnamefont {{Tanvir}}}, \bibinfo {author}
  {\bibfnamefont {A.}~\bibnamefont {{de Ugarte Postigo}}}, \bibinfo {author}
  {\bibfnamefont {C.~D.}\ \bibnamefont {{Kilpatrick}}}, \bibinfo {author}
  {\bibfnamefont {C.~J.}\ \bibnamefont {{Moore}}}, \bibinfo {author}
  {\bibfnamefont {B.~D.}\ \bibnamefont {{Metzger}}}, \bibinfo {author}
  {\bibfnamefont {M.~E.}\ \bibnamefont {{Ravasio}}}, \bibinfo {author}
  {\bibfnamefont {A.}~\bibnamefont {{Rossi}}}, \bibinfo {author} {\bibfnamefont
  {G.}~\bibnamefont {{Schroeder}}}, \bibinfo {author} {\bibfnamefont
  {J.}~\bibnamefont {{Jencson}}}, \bibinfo {author} {\bibfnamefont {D.~J.}\
  \bibnamefont {{Sand}}}, \bibinfo {author} {\bibfnamefont {N.}~\bibnamefont
  {{Smith}}}, \bibinfo {author} {\bibfnamefont {J.~F.}\ \bibnamefont
  {{Ag{\"u}{\'\i} Fern{\'a}ndez}}}, \bibinfo {author} {\bibfnamefont
  {E.}~\bibnamefont {{Berger}}}, \bibinfo {author} {\bibfnamefont {P.~K.}\
  \bibnamefont {{Blanchard}}}, \bibinfo {author} {\bibfnamefont
  {R.}~\bibnamefont {{Chornock}}}, \bibinfo {author} {\bibfnamefont {B.~E.}\
  \bibnamefont {{Cobb}}}, \bibinfo {author} {\bibfnamefont {M.}~\bibnamefont
  {{De Pasquale}}}, \bibinfo {author} {\bibfnamefont {J.~P.~U.}\ \bibnamefont
  {{Fynbo}}}, \bibinfo {author} {\bibfnamefont {L.}~\bibnamefont {{Izzo}}},
  \bibinfo {author} {\bibfnamefont {D.~A.}\ \bibnamefont {{Kann}}}, \bibinfo
  {author} {\bibfnamefont {T.}~\bibnamefont {{Laskar}}}, \bibinfo {author}
  {\bibfnamefont {E.}~\bibnamefont {{Marini}}}, \bibinfo {author}
  {\bibfnamefont {K.}~\bibnamefont {{Paterson}}}, \bibinfo {author}
  {\bibfnamefont {A.~R.}\ \bibnamefont {{Escorial}}}, \bibinfo {author}
  {\bibfnamefont {H.~M.}\ \bibnamefont {{Sears}}}, \ and\ \bibinfo {author}
  {\bibfnamefont {C.~C.}\ \bibnamefont {{Th{\"o}ne}}},\ }\href {\doibase
  10.1038/s41586-022-05390-w} {\bibfield  {journal} {\bibinfo  {journal}
  {\nat}\ }\textbf {\bibinfo {volume} {612}},\ \bibinfo {pages} {223} (\bibinfo
  {year} {2022})},\ \Eprint {http://arxiv.org/abs/2204.10864} {arXiv:2204.10864
  [astro-ph.HE]} \BibitemShut {NoStop}%
\bibitem [{\citenamefont {{Xiao}}\ \emph {et~al.}(2022)\citenamefont {{Xiao}},
  \citenamefont {{Zhang}}, \citenamefont {{Zhu}}, \citenamefont {{Xiong}},
  \citenamefont {{Gao}}, \citenamefont {{Xu}}, \citenamefont {{Zhang}},
  \citenamefont {{Peng}}, \citenamefont {{Li}}, \citenamefont {{Zhang}},
  \citenamefont {{Lu}}, \citenamefont {{Lin}}, \citenamefont {{Liu}},
  \citenamefont {{Zhang}}, \citenamefont {{Ge}}, \citenamefont {{Tuo}},
  \citenamefont {{Xue}}, \citenamefont {{Fu}}, \citenamefont {{Liu}},
  \citenamefont {{Li}}, \citenamefont {{Wang}}, \citenamefont {{Zheng}},
  \citenamefont {{Wang}}, \citenamefont {{Jiang}}, \citenamefont {{Li}},
  \citenamefont {{Liu}}, \citenamefont {{Cao}}, \citenamefont {{Cai}},
  \citenamefont {{Yi}}, \citenamefont {{Zhao}}, \citenamefont {{Xie}},
  \citenamefont {{Li}}, \citenamefont {{Luo}}, \citenamefont {{Liao}},
  \citenamefont {{Song}}, \citenamefont {{Zhang}}, \citenamefont {{Qu}},
  \citenamefont {{Liu}}, \citenamefont {{Li}}, \citenamefont {{Xu}},\ and\
  \citenamefont {{Li}}}]{xiao2022}%
  \BibitemOpen
  \bibfield  {author} {\bibinfo {author} {\bibfnamefont {S.}~\bibnamefont
  {{Xiao}}}, \bibinfo {author} {\bibfnamefont {Y.-Q.}\ \bibnamefont {{Zhang}}},
  \bibinfo {author} {\bibfnamefont {Z.-P.}\ \bibnamefont {{Zhu}}}, \bibinfo
  {author} {\bibfnamefont {S.-L.}\ \bibnamefont {{Xiong}}}, \bibinfo {author}
  {\bibfnamefont {H.}~\bibnamefont {{Gao}}}, \bibinfo {author} {\bibfnamefont
  {D.}~\bibnamefont {{Xu}}}, \bibinfo {author} {\bibfnamefont {S.-N.}\
  \bibnamefont {{Zhang}}}, \bibinfo {author} {\bibfnamefont {W.-X.}\
  \bibnamefont {{Peng}}}, \bibinfo {author} {\bibfnamefont {X.-B.}\
  \bibnamefont {{Li}}}, \bibinfo {author} {\bibfnamefont {P.}~\bibnamefont
  {{Zhang}}}, \bibinfo {author} {\bibfnamefont {F.-J.}\ \bibnamefont {{Lu}}},
  \bibinfo {author} {\bibfnamefont {L.}~\bibnamefont {{Lin}}}, \bibinfo
  {author} {\bibfnamefont {L.-D.}\ \bibnamefont {{Liu}}}, \bibinfo {author}
  {\bibfnamefont {Z.}~\bibnamefont {{Zhang}}}, \bibinfo {author} {\bibfnamefont
  {M.-Y.}\ \bibnamefont {{Ge}}}, \bibinfo {author} {\bibfnamefont {Y.-L.}\
  \bibnamefont {{Tuo}}}, \bibinfo {author} {\bibfnamefont {W.-C.}\ \bibnamefont
  {{Xue}}}, \bibinfo {author} {\bibfnamefont {S.-Y.}\ \bibnamefont {{Fu}}},
  \bibinfo {author} {\bibfnamefont {X.}~\bibnamefont {{Liu}}}, \bibinfo
  {author} {\bibfnamefont {A.}~\bibnamefont {{Li}}}, \bibinfo {author}
  {\bibfnamefont {T.-C.}\ \bibnamefont {{Wang}}}, \bibinfo {author}
  {\bibfnamefont {C.}~\bibnamefont {{Zheng}}}, \bibinfo {author} {\bibfnamefont
  {Y.}~\bibnamefont {{Wang}}}, \bibinfo {author} {\bibfnamefont {S.-Q.}\
  \bibnamefont {{Jiang}}}, \bibinfo {author} {\bibfnamefont {J.-D.}\
  \bibnamefont {{Li}}}, \bibinfo {author} {\bibfnamefont {J.-C.}\ \bibnamefont
  {{Liu}}}, \bibinfo {author} {\bibfnamefont {Z.-J.}\ \bibnamefont {{Cao}}},
  \bibinfo {author} {\bibfnamefont {C.}~\bibnamefont {{Cai}}}, \bibinfo
  {author} {\bibfnamefont {Q.-B.}\ \bibnamefont {{Yi}}}, \bibinfo {author}
  {\bibfnamefont {Y.}~\bibnamefont {{Zhao}}}, \bibinfo {author} {\bibfnamefont
  {S.-L.}\ \bibnamefont {{Xie}}}, \bibinfo {author} {\bibfnamefont {C.-K.}\
  \bibnamefont {{Li}}}, \bibinfo {author} {\bibfnamefont {Q.}~\bibnamefont
  {{Luo}}}, \bibinfo {author} {\bibfnamefont {J.-Y.}\ \bibnamefont {{Liao}}},
  \bibinfo {author} {\bibfnamefont {L.-M.}\ \bibnamefont {{Song}}}, \bibinfo
  {author} {\bibfnamefont {S.}~\bibnamefont {{Zhang}}}, \bibinfo {author}
  {\bibfnamefont {J.-L.}\ \bibnamefont {{Qu}}}, \bibinfo {author}
  {\bibfnamefont {C.-Z.}\ \bibnamefont {{Liu}}}, \bibinfo {author}
  {\bibfnamefont {X.-F.}\ \bibnamefont {{Li}}}, \bibinfo {author}
  {\bibfnamefont {Y.-P.}\ \bibnamefont {{Xu}}}, \ and\ \bibinfo {author}
  {\bibfnamefont {T.-P.}\ \bibnamefont {{Li}}},\ }\href {\doibase
  10.48550/arXiv.2205.02186} {\bibfield  {journal} {\bibinfo  {journal} {arXiv
  e-prints}\ ,\ \bibinfo {eid} {arXiv:2205.02186}} (\bibinfo {year} {2022})},\
  \Eprint {http://arxiv.org/abs/2205.02186} {arXiv:2205.02186 [astro-ph.HE]}
  \BibitemShut {NoStop}%
\bibitem [{\citenamefont {{Carrasco}}\ and\ \citenamefont
  {{Shibata}}(2020)}]{carrasco2020}%
  \BibitemOpen
  \bibfield  {author} {\bibinfo {author} {\bibfnamefont {F.}~\bibnamefont
  {{Carrasco}}}\ and\ \bibinfo {author} {\bibfnamefont {M.}~\bibnamefont
  {{Shibata}}},\ }\href {\doibase 10.1103/PhysRevD.101.063017} {\bibfield
  {journal} {\bibinfo  {journal} {\prd}\ }\textbf {\bibinfo {volume} {101}},\
  \bibinfo {eid} {063017} (\bibinfo {year} {2020})},\ \Eprint
  {http://arxiv.org/abs/2001.04210} {arXiv:2001.04210 [astro-ph.HE]}
  \BibitemShut {NoStop}%
\bibitem [{\citenamefont {{Zhang}}\ \emph {et~al.}(2022)\citenamefont
  {{Zhang}}, \citenamefont {{Yi}}, \citenamefont {{Zhang}}, \citenamefont
  {{Xiong}},\ and\ \citenamefont {{Xiao}}}]{Zhang2022}%
  \BibitemOpen
  \bibfield  {author} {\bibinfo {author} {\bibfnamefont {Z.}~\bibnamefont
  {{Zhang}}}, \bibinfo {author} {\bibfnamefont {S.-X.}\ \bibnamefont {{Yi}}},
  \bibinfo {author} {\bibfnamefont {S.-N.}\ \bibnamefont {{Zhang}}}, \bibinfo
  {author} {\bibfnamefont {S.-L.}\ \bibnamefont {{Xiong}}}, \ and\ \bibinfo
  {author} {\bibfnamefont {S.}~\bibnamefont {{Xiao}}},\ }\href {\doibase
  10.3847/2041-8213/ac9b55} {\bibfield  {journal} {\bibinfo  {journal} {\apjl}\
  }\textbf {\bibinfo {volume} {939}},\ \bibinfo {eid} {L25} (\bibinfo {year}
  {2022})},\ \Eprint {http://arxiv.org/abs/2207.12324} {arXiv:2207.12324
  [astro-ph.HE]} \BibitemShut {NoStop}%
\bibitem [{\citenamefont {{Suvorov}}\ \emph {et~al.}(2022)\citenamefont
  {{Suvorov}}, \citenamefont {{Kuan}},\ and\ \citenamefont
  {{Kokkotas}}}]{suvorov2022}%
  \BibitemOpen
  \bibfield  {author} {\bibinfo {author} {\bibfnamefont {A.~G.}\ \bibnamefont
  {{Suvorov}}}, \bibinfo {author} {\bibfnamefont {H.~J.}\ \bibnamefont
  {{Kuan}}}, \ and\ \bibinfo {author} {\bibfnamefont {K.~D.}\ \bibnamefont
  {{Kokkotas}}},\ }\href {\doibase 10.1051/0004-6361/202244082} {\bibfield
  {journal} {\bibinfo  {journal} {\aap}\ }\textbf {\bibinfo {volume} {664}},\
  \bibinfo {eid} {A177} (\bibinfo {year} {2022})},\ \Eprint
  {http://arxiv.org/abs/2205.11112} {arXiv:2205.11112 [astro-ph.HE]}
  \BibitemShut {NoStop}%
\bibitem [{\citenamefont {{Bildsten}}\ and\ \citenamefont
  {{Cutler}}(1992)}]{Bildsten92}%
  \BibitemOpen
  \bibfield  {author} {\bibinfo {author} {\bibfnamefont {L.}~\bibnamefont
  {{Bildsten}}}\ and\ \bibinfo {author} {\bibfnamefont {C.}~\bibnamefont
  {{Cutler}}},\ }\href {\doibase 10.1086/171983} {\bibfield  {journal}
  {\bibinfo  {journal} {Astrophys. J.}\ }\textbf {\bibinfo {volume} {400}},\
  \bibinfo {pages} {175} (\bibinfo {year} {1992})}\BibitemShut {NoStop}%
\bibitem [{\citenamefont {{Zhou}}\ \emph {et~al.}(2004)\citenamefont {{Zhou}},
  \citenamefont {{Xu}}, \citenamefont {{Wu}},\ and\ \citenamefont
  {{Wang}}}]{zhou2004}%
  \BibitemOpen
  \bibfield  {author} {\bibinfo {author} {\bibfnamefont {A.~Z.}\ \bibnamefont
  {{Zhou}}}, \bibinfo {author} {\bibfnamefont {R.~X.}\ \bibnamefont {{Xu}}},
  \bibinfo {author} {\bibfnamefont {X.~J.}\ \bibnamefont {{Wu}}}, \ and\
  \bibinfo {author} {\bibfnamefont {N.}~\bibnamefont {{Wang}}},\ }\href
  {\doibase 10.1016/j.astropartphys.2004.05.007} {\bibfield  {journal}
  {\bibinfo  {journal} {Astroparticle Physics}\ }\textbf {\bibinfo {volume}
  {22}},\ \bibinfo {pages} {73} (\bibinfo {year} {2004})},\ \Eprint
  {http://arxiv.org/abs/astro-ph/0404554} {arXiv:astro-ph/0404554 [astro-ph]}
  \BibitemShut {NoStop}%
\bibitem [{\citenamefont {{Baym}}\ and\ \citenamefont
  {{Pines}}(1971)}]{Baym71c}%
  \BibitemOpen
  \bibfield  {author} {\bibinfo {author} {\bibfnamefont {G.}~\bibnamefont
  {{Baym}}}\ and\ \bibinfo {author} {\bibfnamefont {D.}~\bibnamefont
  {{Pines}}},\ }\href {\doibase 10.1016/0003-4916(71)90084-4} {\bibfield
  {journal} {\bibinfo  {journal} {Annals of Physics}\ }\textbf {\bibinfo
  {volume} {66}},\ \bibinfo {pages} {816} (\bibinfo {year} {1971})}\BibitemShut
  {NoStop}%
\bibitem [{\citenamefont {{Zhou}}\ \emph {et~al.}(2014)\citenamefont {{Zhou}},
  \citenamefont {{Lu}}, \citenamefont {{Tong}},\ and\ \citenamefont
  {{Xu}}}]{zhou2014}%
  \BibitemOpen
  \bibfield  {author} {\bibinfo {author} {\bibfnamefont {E.~P.}\ \bibnamefont
  {{Zhou}}}, \bibinfo {author} {\bibfnamefont {J.~G.}\ \bibnamefont {{Lu}}},
  \bibinfo {author} {\bibfnamefont {H.}~\bibnamefont {{Tong}}}, \ and\ \bibinfo
  {author} {\bibfnamefont {R.~X.}\ \bibnamefont {{Xu}}},\ }\href {\doibase
  10.1093/mnras/stu1370} {\bibfield  {journal} {\bibinfo  {journal} {Mon. Not.
  R. Astron. Soc.}\ }\textbf {\bibinfo {volume} {443}},\ \bibinfo {pages}
  {2705} (\bibinfo {year} {2014})},\ \Eprint {http://arxiv.org/abs/1404.2793}
  {arXiv:1404.2793 [astro-ph.HE]} \BibitemShut {NoStop}%
\bibitem [{\citenamefont {{Lai}}\ \emph {et~al.}(2018)\citenamefont {{Lai}},
  \citenamefont {{Yun}}, \citenamefont {{Lu}}, \citenamefont {{L{\"u}}},
  \citenamefont {{Wang}},\ and\ \citenamefont {{Xu}}}]{Lai2018c}%
  \BibitemOpen
  \bibfield  {author} {\bibinfo {author} {\bibfnamefont {X.~Y.}\ \bibnamefont
  {{Lai}}}, \bibinfo {author} {\bibfnamefont {C.~A.}\ \bibnamefont {{Yun}}},
  \bibinfo {author} {\bibfnamefont {J.~G.}\ \bibnamefont {{Lu}}}, \bibinfo
  {author} {\bibfnamefont {G.~L.}\ \bibnamefont {{L{\"u}}}}, \bibinfo {author}
  {\bibfnamefont {Z.~J.}\ \bibnamefont {{Wang}}}, \ and\ \bibinfo {author}
  {\bibfnamefont {R.~X.}\ \bibnamefont {{Xu}}},\ }\href {\doibase
  10.1093/mnras/sty474} {\bibfield  {journal} {\bibinfo  {journal} {\mnras}\
  }\textbf {\bibinfo {volume} {476}},\ \bibinfo {pages} {3303} (\bibinfo {year}
  {2018})},\ \Eprint {http://arxiv.org/abs/1707.07471} {arXiv:1707.07471
  [astro-ph.HE]} \BibitemShut {NoStop}%
\bibitem [{\citenamefont {{Kunert}}\ \emph {et~al.}(2023)\citenamefont
  {{Kunert}}, \citenamefont {{Antier}}, \citenamefont {{Nedora}}, \citenamefont
  {{Bulla}}, \citenamefont {{Pang}}, \citenamefont {{Anand}}, \citenamefont
  {{Coughlin}}, \citenamefont {{Tews}}, \citenamefont {{Barnes}}, \citenamefont
  {{Pilloix}}, \citenamefont {{Kiendrebeogo}},\ and\ \citenamefont
  {{Dietrich}}}]{Kunert2023}%
  \BibitemOpen
  \bibfield  {author} {\bibinfo {author} {\bibfnamefont {N.}~\bibnamefont
  {{Kunert}}}, \bibinfo {author} {\bibfnamefont {S.}~\bibnamefont {{Antier}}},
  \bibinfo {author} {\bibfnamefont {V.}~\bibnamefont {{Nedora}}}, \bibinfo
  {author} {\bibfnamefont {M.}~\bibnamefont {{Bulla}}}, \bibinfo {author}
  {\bibfnamefont {P.~T.~H.}\ \bibnamefont {{Pang}}}, \bibinfo {author}
  {\bibfnamefont {S.}~\bibnamefont {{Anand}}}, \bibinfo {author} {\bibfnamefont
  {M.}~\bibnamefont {{Coughlin}}}, \bibinfo {author} {\bibfnamefont
  {I.}~\bibnamefont {{Tews}}}, \bibinfo {author} {\bibfnamefont
  {J.}~\bibnamefont {{Barnes}}}, \bibinfo {author} {\bibfnamefont
  {M.}~\bibnamefont {{Pilloix}}}, \bibinfo {author} {\bibfnamefont
  {W.}~\bibnamefont {{Kiendrebeogo}}}, \ and\ \bibinfo {author} {\bibfnamefont
  {T.}~\bibnamefont {{Dietrich}}},\ }\href {\doibase 10.48550/arXiv.2301.02049}
  {\bibfield  {journal} {\bibinfo  {journal} {arXiv e-prints}\ ,\ \bibinfo
  {eid} {arXiv:2301.02049}} (\bibinfo {year} {2023})},\ \Eprint
  {http://arxiv.org/abs/2301.02049} {arXiv:2301.02049 [astro-ph.HE]}
  \BibitemShut {NoStop}%
\bibitem [{\citenamefont {{Zhu}}\ \emph {et~al.}(2022)\citenamefont {{Zhu}},
  \citenamefont {{Wang}}, \citenamefont {{Sun}}, \citenamefont {{Yang}},
  \citenamefont {{Li}}, \citenamefont {{Hu}}, \citenamefont {{Qin}},\ and\
  \citenamefont {{Wu}}}]{Zhu2022}%
  \BibitemOpen
  \bibfield  {author} {\bibinfo {author} {\bibfnamefont {J.-P.}\ \bibnamefont
  {{Zhu}}}, \bibinfo {author} {\bibfnamefont {X.~I.}\ \bibnamefont {{Wang}}},
  \bibinfo {author} {\bibfnamefont {H.}~\bibnamefont {{Sun}}}, \bibinfo
  {author} {\bibfnamefont {Y.-P.}\ \bibnamefont {{Yang}}}, \bibinfo {author}
  {\bibfnamefont {Z.}~\bibnamefont {{Li}}}, \bibinfo {author} {\bibfnamefont
  {R.-C.}\ \bibnamefont {{Hu}}}, \bibinfo {author} {\bibfnamefont
  {Y.}~\bibnamefont {{Qin}}}, \ and\ \bibinfo {author} {\bibfnamefont
  {S.}~\bibnamefont {{Wu}}},\ }\href {\doibase 10.3847/2041-8213/ac85ad}
  {\bibfield  {journal} {\bibinfo  {journal} {\apjl}\ }\textbf {\bibinfo
  {volume} {936}},\ \bibinfo {eid} {L10} (\bibinfo {year} {2022})},\ \Eprint
  {http://arxiv.org/abs/2207.10470} {arXiv:2207.10470 [astro-ph.HE]}
  \BibitemShut {NoStop}%
\bibitem [{\citenamefont {{Yang}}\ \emph {et~al.}(2022)\citenamefont {{Yang}},
  \citenamefont {{Ai}}, \citenamefont {{Zhang}}, \citenamefont {{Zhang}},
  \citenamefont {{Liu}}, \citenamefont {{Wang}}, \citenamefont {{Yang}},
  \citenamefont {{Yin}}, \citenamefont {{Li}},\ and\ \citenamefont
  {{L{\"u}}}}]{yang2022}%
  \BibitemOpen
  \bibfield  {author} {\bibinfo {author} {\bibfnamefont {J.}~\bibnamefont
  {{Yang}}}, \bibinfo {author} {\bibfnamefont {S.}~\bibnamefont {{Ai}}},
  \bibinfo {author} {\bibfnamefont {B.-B.}\ \bibnamefont {{Zhang}}}, \bibinfo
  {author} {\bibfnamefont {B.}~\bibnamefont {{Zhang}}}, \bibinfo {author}
  {\bibfnamefont {Z.-K.}\ \bibnamefont {{Liu}}}, \bibinfo {author}
  {\bibfnamefont {X.~I.}\ \bibnamefont {{Wang}}}, \bibinfo {author}
  {\bibfnamefont {Y.-H.}\ \bibnamefont {{Yang}}}, \bibinfo {author}
  {\bibfnamefont {Y.-H.}\ \bibnamefont {{Yin}}}, \bibinfo {author}
  {\bibfnamefont {Y.}~\bibnamefont {{Li}}}, \ and\ \bibinfo {author}
  {\bibfnamefont {H.-J.}\ \bibnamefont {{L{\"u}}}},\ }\href {\doibase
  10.1038/s41586-022-05403-8} {\bibfield  {journal} {\bibinfo  {journal}
  {\nat}\ }\textbf {\bibinfo {volume} {612}},\ \bibinfo {pages} {232} (\bibinfo
  {year} {2022})},\ \Eprint {http://arxiv.org/abs/2204.12771} {arXiv:2204.12771
  [astro-ph.HE]} \BibitemShut {NoStop}%
\bibitem [{\citenamefont {{Ferdman}}\ \emph {et~al.}(2020)\citenamefont
  {{Ferdman}}, \citenamefont {{Freire}}, \citenamefont {{Perera}},
  \citenamefont {{Pol}}, \citenamefont {{Camilo}}, \citenamefont
  {{Chatterjee}}, \citenamefont {{Cordes}}, \citenamefont {{Crawford}},
  \citenamefont {{Hessels}}, \citenamefont {{Kaspi}}, \citenamefont
  {{McLaughlin}}, \citenamefont {{Parent}}, \citenamefont {{Stairs}},\ and\
  \citenamefont {{van Leeuwen}}}]{Ferdman2020}%
  \BibitemOpen
  \bibfield  {author} {\bibinfo {author} {\bibfnamefont {R.~D.}\ \bibnamefont
  {{Ferdman}}}, \bibinfo {author} {\bibfnamefont {P.~C.~C.}\ \bibnamefont
  {{Freire}}}, \bibinfo {author} {\bibfnamefont {B.~B.~P.}\ \bibnamefont
  {{Perera}}}, \bibinfo {author} {\bibfnamefont {N.}~\bibnamefont {{Pol}}},
  \bibinfo {author} {\bibfnamefont {F.}~\bibnamefont {{Camilo}}}, \bibinfo
  {author} {\bibfnamefont {S.}~\bibnamefont {{Chatterjee}}}, \bibinfo {author}
  {\bibfnamefont {J.~M.}\ \bibnamefont {{Cordes}}}, \bibinfo {author}
  {\bibfnamefont {F.}~\bibnamefont {{Crawford}}}, \bibinfo {author}
  {\bibfnamefont {J.~W.~T.}\ \bibnamefont {{Hessels}}}, \bibinfo {author}
  {\bibfnamefont {V.~M.}\ \bibnamefont {{Kaspi}}}, \bibinfo {author}
  {\bibfnamefont {M.~A.}\ \bibnamefont {{McLaughlin}}}, \bibinfo {author}
  {\bibfnamefont {E.}~\bibnamefont {{Parent}}}, \bibinfo {author}
  {\bibfnamefont {I.~H.}\ \bibnamefont {{Stairs}}}, \ and\ \bibinfo {author}
  {\bibfnamefont {J.}~\bibnamefont {{van Leeuwen}}},\ }\href {\doibase
  10.1038/s41586-020-2439-x} {\bibfield  {journal} {\bibinfo  {journal} {\nat}\
  }\textbf {\bibinfo {volume} {583}},\ \bibinfo {pages} {211} (\bibinfo {year}
  {2020})},\ \Eprint {http://arxiv.org/abs/2007.04175} {arXiv:2007.04175
  [astro-ph.HE]} \BibitemShut {NoStop}%
\bibitem [{\citenamefont {{The LIGO Scientific Collaboration}}\ \emph
  {et~al.}(2020)\citenamefont {{The LIGO Scientific Collaboration}},
  \citenamefont {{the Virgo Collaboration}}, \citenamefont {{Abbott}},
  \citenamefont {{Abbott}}, \citenamefont {{Abbott}}, \citenamefont
  {{Abraham}}, \citenamefont {{Acernese}}, \citenamefont {{Ackley}},
  \citenamefont {{Adams}}, \citenamefont {{Adhikari}}, \citenamefont {{Adya}},\
  and\ \citenamefont {et~al.}}]{Abbott2020}%
  \BibitemOpen
  \bibfield  {author} {\bibinfo {author} {\bibnamefont {{The LIGO Scientific
  Collaboration}}}, \bibinfo {author} {\bibnamefont {{the Virgo
  Collaboration}}}, \bibinfo {author} {\bibfnamefont {B.~P.}\ \bibnamefont
  {{Abbott}}}, \bibinfo {author} {\bibfnamefont {R.}~\bibnamefont {{Abbott}}},
  \bibinfo {author} {\bibfnamefont {T.~D.}\ \bibnamefont {{Abbott}}}, \bibinfo
  {author} {\bibfnamefont {S.}~\bibnamefont {{Abraham}}}, \bibinfo {author}
  {\bibfnamefont {F.}~\bibnamefont {{Acernese}}}, \bibinfo {author}
  {\bibfnamefont {K.}~\bibnamefont {{Ackley}}}, \bibinfo {author}
  {\bibfnamefont {C.}~\bibnamefont {{Adams}}}, \bibinfo {author} {\bibfnamefont
  {R.~X.}\ \bibnamefont {{Adhikari}}}, \bibinfo {author} {\bibfnamefont
  {V.~B.}\ \bibnamefont {{Adya}}}, \ and\ \bibinfo {author} {\bibnamefont
  {et~al.}},\ }\href@noop {} {\bibfield  {journal} {\bibinfo  {journal} {arXiv
  e-prints}\ ,\ \bibinfo {eid} {arXiv:2001.01761}} (\bibinfo {year} {2020})},\
  \Eprint {http://arxiv.org/abs/2001.01761} {arXiv:2001.01761 [astro-ph.HE]}
  \BibitemShut {NoStop}%
\bibitem [{\citenamefont {{Lai}}\ and\ \citenamefont {{Xu}}(2009)}]{lai2009}%
  \BibitemOpen
  \bibfield  {author} {\bibinfo {author} {\bibfnamefont {X.~Y.}\ \bibnamefont
  {{Lai}}}\ and\ \bibinfo {author} {\bibfnamefont {R.~X.}\ \bibnamefont
  {{Xu}}},\ }\href {\doibase 10.1111/j.1745-3933.2009.00701.x} {\bibfield
  {journal} {\bibinfo  {journal} {Mon. Not. R. Astron. Soc.}\ }\textbf
  {\bibinfo {volume} {398}},\ \bibinfo {pages} {L31} (\bibinfo {year}
  {2009})},\ \Eprint {http://arxiv.org/abs/0905.2839} {arXiv:0905.2839
  [astro-ph.HE]} \BibitemShut {NoStop}%
\bibitem [{\citenamefont {{Kyutoku}}\ \emph {et~al.}(2015)\citenamefont
  {{Kyutoku}}, \citenamefont {{Ioka}}, \citenamefont {{Okawa}}, \citenamefont
  {{Shibata}},\ and\ \citenamefont {{Taniguchi}}}]{Kyutoku2015}%
  \BibitemOpen
  \bibfield  {author} {\bibinfo {author} {\bibfnamefont {K.}~\bibnamefont
  {{Kyutoku}}}, \bibinfo {author} {\bibfnamefont {K.}~\bibnamefont {{Ioka}}},
  \bibinfo {author} {\bibfnamefont {H.}~\bibnamefont {{Okawa}}}, \bibinfo
  {author} {\bibfnamefont {M.}~\bibnamefont {{Shibata}}}, \ and\ \bibinfo
  {author} {\bibfnamefont {K.}~\bibnamefont {{Taniguchi}}},\ }\href {\doibase
  10.1103/PhysRevD.92.044028} {\bibfield  {journal} {\bibinfo  {journal} {Phys.
  Rev. D}\ }\textbf {\bibinfo {volume} {92}},\ \bibinfo {eid} {044028}
  (\bibinfo {year} {2015})},\ \Eprint {http://arxiv.org/abs/1502.05402}
  {arXiv:1502.05402 [astro-ph.HE]} \BibitemShut {NoStop}%
\bibitem [{\citenamefont {{Kawaguchi}}\ \emph {et~al.}(2016)\citenamefont
  {{Kawaguchi}}, \citenamefont {{Kyutoku}}, \citenamefont {{Shibata}},\ and\
  \citenamefont {{Tanaka}}}]{Kawaguchi2016}%
  \BibitemOpen
  \bibfield  {author} {\bibinfo {author} {\bibfnamefont {K.}~\bibnamefont
  {{Kawaguchi}}}, \bibinfo {author} {\bibfnamefont {K.}~\bibnamefont
  {{Kyutoku}}}, \bibinfo {author} {\bibfnamefont {M.}~\bibnamefont
  {{Shibata}}}, \ and\ \bibinfo {author} {\bibfnamefont {M.}~\bibnamefont
  {{Tanaka}}},\ }\href {\doibase 10.3847/0004-637X/825/1/52} {\bibfield
  {journal} {\bibinfo  {journal} {\apj}\ }\textbf {\bibinfo {volume} {825}},\
  \bibinfo {eid} {52} (\bibinfo {year} {2016})},\ \Eprint
  {http://arxiv.org/abs/1601.07711} {arXiv:1601.07711 [astro-ph.HE]}
  \BibitemShut {NoStop}%
\bibitem [{\citenamefont {Riemenschneider}\ \emph {et~al.}(2021)\citenamefont
  {Riemenschneider}, \citenamefont {Rettegno}, \citenamefont {Breschi},
  \citenamefont {Albertini}, \citenamefont {Gamba}, \citenamefont {Bernuzzi},\
  and\ \citenamefont {Nagar}}]{Riemenschneider2021}%
  \BibitemOpen
  \bibfield  {author} {\bibinfo {author} {\bibfnamefont {G.}~\bibnamefont
  {Riemenschneider}}, \bibinfo {author} {\bibfnamefont {P.}~\bibnamefont
  {Rettegno}}, \bibinfo {author} {\bibfnamefont {M.}~\bibnamefont {Breschi}},
  \bibinfo {author} {\bibfnamefont {A.}~\bibnamefont {Albertini}}, \bibinfo
  {author} {\bibfnamefont {R.}~\bibnamefont {Gamba}}, \bibinfo {author}
  {\bibfnamefont {S.}~\bibnamefont {Bernuzzi}}, \ and\ \bibinfo {author}
  {\bibfnamefont {A.}~\bibnamefont {Nagar}},\ }\href {\doibase
  10.1103/PhysRevD.104.104045} {\bibfield  {journal} {\bibinfo  {journal}
  {Phys. Rev. D}\ }\textbf {\bibinfo {volume} {104}},\ \bibinfo {pages}
  {104045} (\bibinfo {year} {2021})},\ \Eprint
  {http://arxiv.org/abs/2104.07533} {arXiv:2104.07533 [gr-qc]} \BibitemShut
  {NoStop}%
\bibitem [{\citenamefont {Nagar}\ \emph
  {et~al.}(2020{\natexlab{a}})\citenamefont {Nagar}, \citenamefont
  {Riemenschneider}, \citenamefont {Pratten}, \citenamefont {Rettegno},\ and\
  \citenamefont {Messina}}]{Nagar2020}%
  \BibitemOpen
  \bibfield  {author} {\bibinfo {author} {\bibfnamefont {A.}~\bibnamefont
  {Nagar}}, \bibinfo {author} {\bibfnamefont {G.}~\bibnamefont
  {Riemenschneider}}, \bibinfo {author} {\bibfnamefont {G.}~\bibnamefont
  {Pratten}}, \bibinfo {author} {\bibfnamefont {P.}~\bibnamefont {Rettegno}}, \
  and\ \bibinfo {author} {\bibfnamefont {F.}~\bibnamefont {Messina}},\ }\href
  {\doibase 10.1103/PhysRevD.102.024077} {\bibfield  {journal} {\bibinfo
  {journal} {Phys. Rev. D}\ }\textbf {\bibinfo {volume} {102}},\ \bibinfo
  {pages} {024077} (\bibinfo {year} {2020}{\natexlab{a}})},\ \Eprint
  {http://arxiv.org/abs/2001.09082} {arXiv:2001.09082 [gr-qc]} \BibitemShut
  {NoStop}%
\bibitem [{\citenamefont {Nagar}\ \emph
  {et~al.}(2020{\natexlab{b}})\citenamefont {Nagar}, \citenamefont {Pratten},
  \citenamefont {Riemenschneider},\ and\ \citenamefont {Gamba}}]{Nagar2019}%
  \BibitemOpen
  \bibfield  {author} {\bibinfo {author} {\bibfnamefont {A.}~\bibnamefont
  {Nagar}}, \bibinfo {author} {\bibfnamefont {G.}~\bibnamefont {Pratten}},
  \bibinfo {author} {\bibfnamefont {G.}~\bibnamefont {Riemenschneider}}, \ and\
  \bibinfo {author} {\bibfnamefont {R.}~\bibnamefont {Gamba}},\ }\href
  {\doibase 10.1103/PhysRevD.101.024041} {\bibfield  {journal} {\bibinfo
  {journal} {Phys. Rev. D}\ }\textbf {\bibinfo {volume} {101}},\ \bibinfo
  {pages} {024041} (\bibinfo {year} {2020}{\natexlab{b}})},\ \Eprint
  {http://arxiv.org/abs/1904.09550} {arXiv:1904.09550 [gr-qc]} \BibitemShut
  {NoStop}%
\bibitem [{\citenamefont {Nagar}\ \emph {et~al.}(2018)\citenamefont {Nagar}
  \emph {et~al.}}]{Nagar2018}%
  \BibitemOpen
  \bibfield  {author} {\bibinfo {author} {\bibfnamefont {A.}~\bibnamefont
  {Nagar}} \emph {et~al.},\ }\href {\doibase 10.1103/PhysRevD.98.104052}
  {\bibfield  {journal} {\bibinfo  {journal} {Phys. Rev. D}\ }\textbf {\bibinfo
  {volume} {98}},\ \bibinfo {pages} {104052} (\bibinfo {year} {2018})},\
  \Eprint {http://arxiv.org/abs/1806.01772} {arXiv:1806.01772 [gr-qc]}
  \BibitemShut {NoStop}%
\bibitem [{\citenamefont {Nagar}\ \emph {et~al.}(2016)\citenamefont {Nagar},
  \citenamefont {Damour}, \citenamefont {Reisswig},\ and\ \citenamefont
  {Pollney}}]{Nagar2015}%
  \BibitemOpen
  \bibfield  {author} {\bibinfo {author} {\bibfnamefont {A.}~\bibnamefont
  {Nagar}}, \bibinfo {author} {\bibfnamefont {T.}~\bibnamefont {Damour}},
  \bibinfo {author} {\bibfnamefont {C.}~\bibnamefont {Reisswig}}, \ and\
  \bibinfo {author} {\bibfnamefont {D.}~\bibnamefont {Pollney}},\ }\href
  {\doibase 10.1103/PhysRevD.93.044046} {\bibfield  {journal} {\bibinfo
  {journal} {Phys. Rev. D}\ }\textbf {\bibinfo {volume} {93}},\ \bibinfo
  {pages} {044046} (\bibinfo {year} {2016})},\ \Eprint
  {http://arxiv.org/abs/1506.08457} {arXiv:1506.08457 [gr-qc]} \BibitemShut
  {NoStop}%
\bibitem [{\citenamefont {Damour}\ and\ \citenamefont
  {Nagar}(2014)}]{Damour2014}%
  \BibitemOpen
  \bibfield  {author} {\bibinfo {author} {\bibfnamefont {T.}~\bibnamefont
  {Damour}}\ and\ \bibinfo {author} {\bibfnamefont {A.}~\bibnamefont {Nagar}},\
  }\href {\doibase 10.1103/PhysRevD.90.044018} {\bibfield  {journal} {\bibinfo
  {journal} {Phys. Rev. D}\ }\textbf {\bibinfo {volume} {90}},\ \bibinfo
  {pages} {044018} (\bibinfo {year} {2014})},\ \Eprint
  {http://arxiv.org/abs/1406.6913} {arXiv:1406.6913 [gr-qc]} \BibitemShut
  {NoStop}%
\bibitem [{\citenamefont {Duncan}(1998)}]{Duncan:1998my}%
  \BibitemOpen
  \bibfield  {author} {\bibinfo {author} {\bibfnamefont {R.~C.}\ \bibnamefont
  {Duncan}},\ }\href {\doibase 10.1086/311303} {\bibfield  {journal} {\bibinfo
  {journal} {Astrophys. J. Lett.}\ }\textbf {\bibinfo {volume} {498}},\
  \bibinfo {pages} {L45} (\bibinfo {year} {1998})},\ \Eprint
  {http://arxiv.org/abs/astro-ph/9803060} {arXiv:astro-ph/9803060} \BibitemShut
  {NoStop}%
\bibitem [{\citenamefont {Watts}\ and\ \citenamefont
  {Strohmayer}(2007)}]{Watts:2006mr}%
  \BibitemOpen
  \bibfield  {author} {\bibinfo {author} {\bibfnamefont {A.~L.}\ \bibnamefont
  {Watts}}\ and\ \bibinfo {author} {\bibfnamefont {T.~E.}\ \bibnamefont
  {Strohmayer}},\ }\href {\doibase 10.1016/j.asr.2006.12.021} {\bibfield
  {journal} {\bibinfo  {journal} {Adv. Space Res.}\ }\textbf {\bibinfo {volume}
  {40}},\ \bibinfo {pages} {1446} (\bibinfo {year} {2007})},\ \Eprint
  {http://arxiv.org/abs/astro-ph/0612252} {arXiv:astro-ph/0612252} \BibitemShut
  {NoStop}%
\bibitem [{\citenamefont {Benioff}\ \emph {et~al.}(1961)\citenamefont
  {Benioff}, \citenamefont {Press},\ and\ \citenamefont {Smith}}]{Benioff1961}%
  \BibitemOpen
  \bibfield  {author} {\bibinfo {author} {\bibfnamefont {H.}~\bibnamefont
  {Benioff}}, \bibinfo {author} {\bibfnamefont {F.}~\bibnamefont {Press}}, \
  and\ \bibinfo {author} {\bibfnamefont {S.}~\bibnamefont {Smith}},\ }\href
  {\doibase https://doi.org/10.1029/JZ066i002p00605} {\bibfield  {journal}
  {\bibinfo  {journal} {Journal of Geophysical Research (1896-1977)}\ }\textbf
  {\bibinfo {volume} {66}},\ \bibinfo {pages} {605} (\bibinfo {year}
  {1961})}\BibitemShut {NoStop}%
\bibitem [{\citenamefont {{Park}}\ \emph {et~al.}(2005)\citenamefont {{Park}},
  \citenamefont {{Song}}, \citenamefont {{Tromp}}, \citenamefont {{Okal}},
  \citenamefont {{Stein}}, \citenamefont {{Roult}}, \citenamefont {{Clevede}},
  \citenamefont {{Laske}}, \citenamefont {{Kanamori}}, \citenamefont {{Davis}},
  \citenamefont {{Berger}}, \citenamefont {{Braitenberg}}, \citenamefont {{Van
  Camp}}, \citenamefont {{Lei}}, \citenamefont {{Sun}}, \citenamefont {{Xu}},\
  and\ \citenamefont {{Rosat}}}]{Park2004}%
  \BibitemOpen
  \bibfield  {author} {\bibinfo {author} {\bibfnamefont {J.}~\bibnamefont
  {{Park}}}, \bibinfo {author} {\bibfnamefont {T.-R.~A.}\ \bibnamefont
  {{Song}}}, \bibinfo {author} {\bibfnamefont {J.}~\bibnamefont {{Tromp}}},
  \bibinfo {author} {\bibfnamefont {E.}~\bibnamefont {{Okal}}}, \bibinfo
  {author} {\bibfnamefont {S.}~\bibnamefont {{Stein}}}, \bibinfo {author}
  {\bibfnamefont {G.}~\bibnamefont {{Roult}}}, \bibinfo {author} {\bibfnamefont
  {E.}~\bibnamefont {{Clevede}}}, \bibinfo {author} {\bibfnamefont
  {G.}~\bibnamefont {{Laske}}}, \bibinfo {author} {\bibfnamefont
  {H.}~\bibnamefont {{Kanamori}}}, \bibinfo {author} {\bibfnamefont
  {P.}~\bibnamefont {{Davis}}}, \bibinfo {author} {\bibfnamefont
  {J.}~\bibnamefont {{Berger}}}, \bibinfo {author} {\bibfnamefont
  {C.}~\bibnamefont {{Braitenberg}}}, \bibinfo {author} {\bibfnamefont
  {M.}~\bibnamefont {{Van Camp}}}, \bibinfo {author} {\bibfnamefont
  {X.}~\bibnamefont {{Lei}}}, \bibinfo {author} {\bibfnamefont
  {H.}~\bibnamefont {{Sun}}}, \bibinfo {author} {\bibfnamefont
  {H.}~\bibnamefont {{Xu}}}, \ and\ \bibinfo {author} {\bibfnamefont
  {S.}~\bibnamefont {{Rosat}}},\ }\href {\doibase 10.1126/science.1112305}
  {\bibfield  {journal} {\bibinfo  {journal} {Science}\ }\textbf {\bibinfo
  {volume} {308}},\ \bibinfo {pages} {1139} (\bibinfo {year}
  {2005})}\BibitemShut {NoStop}%
\bibitem [{\citenamefont {{McDermott}}\ \emph {et~al.}(1988)\citenamefont
  {{McDermott}}, \citenamefont {{van Horn}},\ and\ \citenamefont
  {{Hansen}}}]{McDermott1988}%
  \BibitemOpen
  \bibfield  {author} {\bibinfo {author} {\bibfnamefont {P.~N.}\ \bibnamefont
  {{McDermott}}}, \bibinfo {author} {\bibfnamefont {H.~M.}\ \bibnamefont {{van
  Horn}}}, \ and\ \bibinfo {author} {\bibfnamefont {C.~J.}\ \bibnamefont
  {{Hansen}}},\ }\href {\doibase 10.1086/166044} {\bibfield  {journal}
  {\bibinfo  {journal} {Astrophys. J.}\ }\textbf {\bibinfo {volume} {325}},\
  \bibinfo {pages} {725} (\bibinfo {year} {1988})}\BibitemShut {NoStop}%
\bibitem [{\citenamefont {Lamb}(1881)}]{Lamb1881}%
  \BibitemOpen
  \bibfield  {author} {\bibinfo {author} {\bibfnamefont {H.}~\bibnamefont
  {Lamb}},\ }\href {\doibase https://doi.org/10.1112/plms/s1-13.1.189}
  {\bibfield  {journal} {\bibinfo  {journal} {Proceedings of the London
  Mathematical Society}\ }\textbf {\bibinfo {volume} {s1-13}},\ \bibinfo
  {pages} {189} (\bibinfo {year} {1881})},\ \Eprint
  {http://arxiv.org/abs/https://londmathsoc.onlinelibrary.wiley.com/doi/pdf/10.1112/plms/s1-13.1.189}
  {https://londmathsoc.onlinelibrary.wiley.com/doi/pdf/10.1112/plms/s1-13.1.189}
  \BibitemShut {NoStop}%
\bibitem [{\citenamefont {{Crossley}}(1975)}]{Crossley1975}%
  \BibitemOpen
  \bibfield  {author} {\bibinfo {author} {\bibfnamefont {D.~J.}\ \bibnamefont
  {{Crossley}}},\ }\href@noop {} {\bibfield  {journal} {\bibinfo  {journal}
  {Geophysical Journal}\ }\textbf {\bibinfo {volume} {41}},\ \bibinfo {pages}
  {153} (\bibinfo {year} {1975})}\BibitemShut {NoStop}%
\bibitem [{\citenamefont {{Alterman}}\ \emph {et~al.}(1959)\citenamefont
  {{Alterman}}, \citenamefont {{Jarosch}},\ and\ \citenamefont
  {{Pekeris}}}]{Alterman1959}%
  \BibitemOpen
  \bibfield  {author} {\bibinfo {author} {\bibfnamefont {Z.}~\bibnamefont
  {{Alterman}}}, \bibinfo {author} {\bibfnamefont {H.}~\bibnamefont
  {{Jarosch}}}, \ and\ \bibinfo {author} {\bibfnamefont {C.~L.}\ \bibnamefont
  {{Pekeris}}},\ }\href@noop {} {\bibfield  {journal} {\bibinfo  {journal}
  {Proceedings of the Royal Society of London Series A}\ }\textbf {\bibinfo
  {volume} {252}},\ \bibinfo {pages} {80} (\bibinfo {year} {1959})}\BibitemShut
  {NoStop}%
\bibitem [{\citenamefont {{Li}}\ \emph {et~al.}(2020)\citenamefont {{Li}},
  \citenamefont {{Wen}}, \citenamefont {{Sun}}, \citenamefont {{Liu}},
  \citenamefont {{Liang}}, \citenamefont {{Guo}}, \citenamefont {{Peng}},
  \citenamefont {{Gong}}, \citenamefont {{Li}}, \citenamefont {{Wang}},
  \citenamefont {{Xiong}}, \citenamefont {{Liao}}, \citenamefont {{Lu}},
  \citenamefont {{Wang}}, \citenamefont {{An}}, \citenamefont {{Zhang}},
  \citenamefont {{Gao}}, \citenamefont {{Chen}}, \citenamefont {{Liu}},
  \citenamefont {{Yang}}, \citenamefont {{Qiao}}, \citenamefont {{Zhang}},
  \citenamefont {{Zhao}}, \citenamefont {{Xu}}, \citenamefont {{Zhu}},\ and\
  \citenamefont {{Li}}}]{2020SSPMA..50l9508L}%
  \BibitemOpen
  \bibfield  {author} {\bibinfo {author} {\bibfnamefont {Y.}~\bibnamefont
  {{Li}}}, \bibinfo {author} {\bibfnamefont {X.}~\bibnamefont {{Wen}}},
  \bibinfo {author} {\bibfnamefont {X.}~\bibnamefont {{Sun}}}, \bibinfo
  {author} {\bibfnamefont {X.}~\bibnamefont {{Liu}}}, \bibinfo {author}
  {\bibfnamefont {X.}~\bibnamefont {{Liang}}}, \bibinfo {author} {\bibfnamefont
  {D.}~\bibnamefont {{Guo}}}, \bibinfo {author} {\bibfnamefont
  {W.}~\bibnamefont {{Peng}}}, \bibinfo {author} {\bibfnamefont
  {K.}~\bibnamefont {{Gong}}}, \bibinfo {author} {\bibfnamefont
  {G.}~\bibnamefont {{Li}}}, \bibinfo {author} {\bibfnamefont {H.}~\bibnamefont
  {{Wang}}}, \bibinfo {author} {\bibfnamefont {S.}~\bibnamefont {{Xiong}}},
  \bibinfo {author} {\bibfnamefont {J.}~\bibnamefont {{Liao}}}, \bibinfo
  {author} {\bibfnamefont {H.}~\bibnamefont {{Lu}}}, \bibinfo {author}
  {\bibfnamefont {J.}~\bibnamefont {{Wang}}}, \bibinfo {author} {\bibfnamefont
  {Z.}~\bibnamefont {{An}}}, \bibinfo {author} {\bibfnamefont {D.}~\bibnamefont
  {{Zhang}}}, \bibinfo {author} {\bibfnamefont {M.}~\bibnamefont {{Gao}}},
  \bibinfo {author} {\bibfnamefont {G.}~\bibnamefont {{Chen}}}, \bibinfo
  {author} {\bibfnamefont {Y.}~\bibnamefont {{Liu}}}, \bibinfo {author}
  {\bibfnamefont {S.}~\bibnamefont {{Yang}}}, \bibinfo {author} {\bibfnamefont
  {R.}~\bibnamefont {{Qiao}}}, \bibinfo {author} {\bibfnamefont
  {F.}~\bibnamefont {{Zhang}}}, \bibinfo {author} {\bibfnamefont
  {X.}~\bibnamefont {{Zhao}}}, \bibinfo {author} {\bibfnamefont
  {Y.}~\bibnamefont {{Xu}}}, \bibinfo {author} {\bibfnamefont {Y.}~\bibnamefont
  {{Zhu}}}, \ and\ \bibinfo {author} {\bibfnamefont {X.}~\bibnamefont {{Li}}},\
  }\href {\doibase 10.1360/SSPMA-2019-0417} {\bibfield  {journal} {\bibinfo
  {journal} {Scientia Sinica Physica, Mechanica \& Astronomica}\ }\textbf
  {\bibinfo {volume} {50}},\ \bibinfo {pages} {129508} (\bibinfo {year}
  {2020})}\BibitemShut {NoStop}%
\bibitem [{\citenamefont {{Li}}\ \emph {et~al.}(2018)\citenamefont {{Li}},
  \citenamefont {{Xiong}}, \citenamefont {{Zhang}}, \citenamefont {{Lu}},
  \citenamefont {{Song}}, \citenamefont {{Cao}}, \citenamefont {{Chang}},
  \citenamefont {{Chen}}, \citenamefont {{Chen}}, \citenamefont {{Chen}},
  \citenamefont {{Chen}}, \citenamefont {{Chen}}, \citenamefont {{Chen}},
  \citenamefont {{Cui}}, \citenamefont {{Cui}}, \citenamefont {{Deng}},
  \citenamefont {{Dong}}, \citenamefont {{Du}}, \citenamefont {{Fu}},
  \citenamefont {{Gao}}, \citenamefont {{Gao}}, \citenamefont {{Gao}},
  \citenamefont {{Ge}}, \citenamefont {{Gu}}, \citenamefont {{Guan}},
  \citenamefont {{Guo}}, \citenamefont {{Han}}, \citenamefont {{Hu}},
  \citenamefont {{Huang}}, \citenamefont {{Huo}}, \citenamefont {{Jia}},
  \citenamefont {{Jiang}}, \citenamefont {{Jiang}}, \citenamefont {{Jin}},
  \citenamefont {{Jin}}, \citenamefont {{Li}}, \citenamefont {{Li}},
  \citenamefont {{Li}}, \citenamefont {{Li}}, \citenamefont {{Li}},
  \citenamefont {{Li}}, \citenamefont {{Li}}, \citenamefont {{Li}},
  \citenamefont {{Li}}, \citenamefont {{Li}}, \citenamefont {{Li}},
  \citenamefont {{Liang}}, \citenamefont {{Liao}}, \citenamefont {{Liu}},
  \citenamefont {{Liu}}, \citenamefont {{Liu}}, \citenamefont {{Liu}},
  \citenamefont {{Liu}}, \citenamefont {{Liu}}, \citenamefont {{Liu}},
  \citenamefont {{Lu}}, \citenamefont {{Lu}}, \citenamefont {{Luo}},
  \citenamefont {{Ma}}, \citenamefont {{Meng}}, \citenamefont {{Nang}},
  \citenamefont {{Nie}}, \citenamefont {{Ou}}, \citenamefont {{Qu}},
  \citenamefont {{Sai}}, \citenamefont {{Sun}}, \citenamefont {{Tan}},
  \citenamefont {{Tao}}, \citenamefont {{Tao}}, \citenamefont {{Tuo}},
  \citenamefont {{Wang}}, \citenamefont {{Wang}}, \citenamefont {{Wang}},
  \citenamefont {{Wang}}, \citenamefont {{Wang}}, \citenamefont {{Wen}},
  \citenamefont {{Wu}}, \citenamefont {{Wu}}, \citenamefont {{Xiao}},
  \citenamefont {{Xu}}, \citenamefont {{Xu}}, \citenamefont {{Yan}},
  \citenamefont {{Yang}}, \citenamefont {{Yang}}, \citenamefont {{Yang}},
  \citenamefont {{Zhang}}, \citenamefont {{Zhang}}, \citenamefont {{Zhang}},
  \citenamefont {{Zhang}}, \citenamefont {{Zhang}}, \citenamefont {{Zhang}},
  \citenamefont {{Zhang}}, \citenamefont {{Zhang}}, \citenamefont {{Zhang}},
  \citenamefont {{Zhang}}, \citenamefont {{Zhang}}, \citenamefont {{Zhang}},
  \citenamefont {{Zhang}}, \citenamefont {{Zhang}}, \citenamefont {{Zhang}},
  \citenamefont {{Zhang}}, \citenamefont {{Zhang}}, \citenamefont {{Zhang}},
  \citenamefont {{Zhao}}, \citenamefont {{Zhao}}, \citenamefont {{Zhao}},
  \citenamefont {{Zheng}}, \citenamefont {{Zhu}}, \citenamefont {{Zhu}},\ and\
  \citenamefont {{Zou}}}]{2018SCPMA..61c1011L}%
  \BibitemOpen
  \bibfield  {author} {\bibinfo {author} {\bibfnamefont {T.}~\bibnamefont
  {{Li}}}, \bibinfo {author} {\bibfnamefont {S.}~\bibnamefont {{Xiong}}},
  \bibinfo {author} {\bibfnamefont {S.}~\bibnamefont {{Zhang}}}, \bibinfo
  {author} {\bibfnamefont {F.}~\bibnamefont {{Lu}}}, \bibinfo {author}
  {\bibfnamefont {L.}~\bibnamefont {{Song}}}, \bibinfo {author} {\bibfnamefont
  {X.}~\bibnamefont {{Cao}}}, \bibinfo {author} {\bibfnamefont
  {Z.}~\bibnamefont {{Chang}}}, \bibinfo {author} {\bibfnamefont
  {G.}~\bibnamefont {{Chen}}}, \bibinfo {author} {\bibfnamefont
  {L.}~\bibnamefont {{Chen}}}, \bibinfo {author} {\bibfnamefont
  {T.}~\bibnamefont {{Chen}}}, \bibinfo {author} {\bibfnamefont
  {Y.}~\bibnamefont {{Chen}}}, \bibinfo {author} {\bibfnamefont
  {Y.}~\bibnamefont {{Chen}}}, \bibinfo {author} {\bibfnamefont
  {Y.}~\bibnamefont {{Chen}}}, \bibinfo {author} {\bibfnamefont
  {W.}~\bibnamefont {{Cui}}}, \bibinfo {author} {\bibfnamefont
  {W.}~\bibnamefont {{Cui}}}, \bibinfo {author} {\bibfnamefont
  {J.}~\bibnamefont {{Deng}}}, \bibinfo {author} {\bibfnamefont
  {Y.}~\bibnamefont {{Dong}}}, \bibinfo {author} {\bibfnamefont
  {Y.}~\bibnamefont {{Du}}}, \bibinfo {author} {\bibfnamefont {M.}~\bibnamefont
  {{Fu}}}, \bibinfo {author} {\bibfnamefont {G.}~\bibnamefont {{Gao}}},
  \bibinfo {author} {\bibfnamefont {H.}~\bibnamefont {{Gao}}}, \bibinfo
  {author} {\bibfnamefont {M.}~\bibnamefont {{Gao}}}, \bibinfo {author}
  {\bibfnamefont {M.}~\bibnamefont {{Ge}}}, \bibinfo {author} {\bibfnamefont
  {Y.}~\bibnamefont {{Gu}}}, \bibinfo {author} {\bibfnamefont {J.}~\bibnamefont
  {{Guan}}}, \bibinfo {author} {\bibfnamefont {C.}~\bibnamefont {{Guo}}},
  \bibinfo {author} {\bibfnamefont {D.}~\bibnamefont {{Han}}}, \bibinfo
  {author} {\bibfnamefont {W.}~\bibnamefont {{Hu}}}, \bibinfo {author}
  {\bibfnamefont {Y.}~\bibnamefont {{Huang}}}, \bibinfo {author} {\bibfnamefont
  {J.}~\bibnamefont {{Huo}}}, \bibinfo {author} {\bibfnamefont
  {S.}~\bibnamefont {{Jia}}}, \bibinfo {author} {\bibfnamefont
  {L.}~\bibnamefont {{Jiang}}}, \bibinfo {author} {\bibfnamefont
  {W.}~\bibnamefont {{Jiang}}}, \bibinfo {author} {\bibfnamefont
  {J.}~\bibnamefont {{Jin}}}, \bibinfo {author} {\bibfnamefont
  {Y.}~\bibnamefont {{Jin}}}, \bibinfo {author} {\bibfnamefont
  {B.}~\bibnamefont {{Li}}}, \bibinfo {author} {\bibfnamefont {C.}~\bibnamefont
  {{Li}}}, \bibinfo {author} {\bibfnamefont {G.}~\bibnamefont {{Li}}}, \bibinfo
  {author} {\bibfnamefont {M.}~\bibnamefont {{Li}}}, \bibinfo {author}
  {\bibfnamefont {W.}~\bibnamefont {{Li}}}, \bibinfo {author} {\bibfnamefont
  {X.}~\bibnamefont {{Li}}}, \bibinfo {author} {\bibfnamefont {X.}~\bibnamefont
  {{Li}}}, \bibinfo {author} {\bibfnamefont {X.}~\bibnamefont {{Li}}}, \bibinfo
  {author} {\bibfnamefont {Y.}~\bibnamefont {{Li}}}, \bibinfo {author}
  {\bibfnamefont {Z.}~\bibnamefont {{Li}}}, \bibinfo {author} {\bibfnamefont
  {Z.}~\bibnamefont {{Li}}}, \bibinfo {author} {\bibfnamefont {X.}~\bibnamefont
  {{Liang}}}, \bibinfo {author} {\bibfnamefont {J.}~\bibnamefont {{Liao}}},
  \bibinfo {author} {\bibfnamefont {C.}~\bibnamefont {{Liu}}}, \bibinfo
  {author} {\bibfnamefont {G.}~\bibnamefont {{Liu}}}, \bibinfo {author}
  {\bibfnamefont {H.}~\bibnamefont {{Liu}}}, \bibinfo {author} {\bibfnamefont
  {S.}~\bibnamefont {{Liu}}}, \bibinfo {author} {\bibfnamefont
  {X.}~\bibnamefont {{Liu}}}, \bibinfo {author} {\bibfnamefont
  {Y.}~\bibnamefont {{Liu}}}, \bibinfo {author} {\bibfnamefont
  {Y.}~\bibnamefont {{Liu}}}, \bibinfo {author} {\bibfnamefont
  {B.}~\bibnamefont {{Lu}}}, \bibinfo {author} {\bibfnamefont {X.}~\bibnamefont
  {{Lu}}}, \bibinfo {author} {\bibfnamefont {T.}~\bibnamefont {{Luo}}},
  \bibinfo {author} {\bibfnamefont {X.}~\bibnamefont {{Ma}}}, \bibinfo {author}
  {\bibfnamefont {B.}~\bibnamefont {{Meng}}}, \bibinfo {author} {\bibfnamefont
  {Y.}~\bibnamefont {{Nang}}}, \bibinfo {author} {\bibfnamefont
  {J.}~\bibnamefont {{Nie}}}, \bibinfo {author} {\bibfnamefont
  {G.}~\bibnamefont {{Ou}}}, \bibinfo {author} {\bibfnamefont {J.}~\bibnamefont
  {{Qu}}}, \bibinfo {author} {\bibfnamefont {N.}~\bibnamefont {{Sai}}},
  \bibinfo {author} {\bibfnamefont {L.}~\bibnamefont {{Sun}}}, \bibinfo
  {author} {\bibfnamefont {Y.}~\bibnamefont {{Tan}}}, \bibinfo {author}
  {\bibfnamefont {L.}~\bibnamefont {{Tao}}}, \bibinfo {author} {\bibfnamefont
  {W.}~\bibnamefont {{Tao}}}, \bibinfo {author} {\bibfnamefont
  {Y.}~\bibnamefont {{Tuo}}}, \bibinfo {author} {\bibfnamefont
  {G.}~\bibnamefont {{Wang}}}, \bibinfo {author} {\bibfnamefont
  {H.}~\bibnamefont {{Wang}}}, \bibinfo {author} {\bibfnamefont
  {J.}~\bibnamefont {{Wang}}}, \bibinfo {author} {\bibfnamefont
  {W.}~\bibnamefont {{Wang}}}, \bibinfo {author} {\bibfnamefont
  {Y.}~\bibnamefont {{Wang}}}, \bibinfo {author} {\bibfnamefont
  {X.}~\bibnamefont {{Wen}}}, \bibinfo {author} {\bibfnamefont
  {B.}~\bibnamefont {{Wu}}}, \bibinfo {author} {\bibfnamefont {M.}~\bibnamefont
  {{Wu}}}, \bibinfo {author} {\bibfnamefont {G.}~\bibnamefont {{Xiao}}},
  \bibinfo {author} {\bibfnamefont {H.}~\bibnamefont {{Xu}}}, \bibinfo {author}
  {\bibfnamefont {Y.}~\bibnamefont {{Xu}}}, \bibinfo {author} {\bibfnamefont
  {L.}~\bibnamefont {{Yan}}}, \bibinfo {author} {\bibfnamefont
  {J.}~\bibnamefont {{Yang}}}, \bibinfo {author} {\bibfnamefont
  {S.}~\bibnamefont {{Yang}}}, \bibinfo {author} {\bibfnamefont
  {Y.}~\bibnamefont {{Yang}}}, \bibinfo {author} {\bibfnamefont
  {A.}~\bibnamefont {{Zhang}}}, \bibinfo {author} {\bibfnamefont
  {C.}~\bibnamefont {{Zhang}}}, \bibinfo {author} {\bibfnamefont
  {C.}~\bibnamefont {{Zhang}}}, \bibinfo {author} {\bibfnamefont
  {F.}~\bibnamefont {{Zhang}}}, \bibinfo {author} {\bibfnamefont
  {H.}~\bibnamefont {{Zhang}}}, \bibinfo {author} {\bibfnamefont
  {J.}~\bibnamefont {{Zhang}}}, \bibinfo {author} {\bibfnamefont
  {Q.}~\bibnamefont {{Zhang}}}, \bibinfo {author} {\bibfnamefont
  {S.}~\bibnamefont {{Zhang}}}, \bibinfo {author} {\bibfnamefont
  {T.}~\bibnamefont {{Zhang}}}, \bibinfo {author} {\bibfnamefont
  {W.}~\bibnamefont {{Zhang}}}, \bibinfo {author} {\bibfnamefont
  {W.}~\bibnamefont {{Zhang}}}, \bibinfo {author} {\bibfnamefont
  {W.}~\bibnamefont {{Zhang}}}, \bibinfo {author} {\bibfnamefont
  {Y.}~\bibnamefont {{Zhang}}}, \bibinfo {author} {\bibfnamefont
  {Y.}~\bibnamefont {{Zhang}}}, \bibinfo {author} {\bibfnamefont
  {Y.}~\bibnamefont {{Zhang}}}, \bibinfo {author} {\bibfnamefont
  {Y.}~\bibnamefont {{Zhang}}}, \bibinfo {author} {\bibfnamefont
  {Z.}~\bibnamefont {{Zhang}}}, \bibinfo {author} {\bibfnamefont
  {Z.}~\bibnamefont {{Zhang}}}, \bibinfo {author} {\bibfnamefont
  {H.}~\bibnamefont {{Zhao}}}, \bibinfo {author} {\bibfnamefont
  {J.}~\bibnamefont {{Zhao}}}, \bibinfo {author} {\bibfnamefont
  {X.}~\bibnamefont {{Zhao}}}, \bibinfo {author} {\bibfnamefont
  {S.}~\bibnamefont {{Zheng}}}, \bibinfo {author} {\bibfnamefont
  {Y.}~\bibnamefont {{Zhu}}}, \bibinfo {author} {\bibfnamefont
  {Y.}~\bibnamefont {{Zhu}}}, \ and\ \bibinfo {author} {\bibfnamefont
  {C.}~\bibnamefont {{Zou}}},\ }\href {\doibase 10.1007/s11433-017-9107-5}
  {\bibfield  {journal} {\bibinfo  {journal} {Science China Physics, Mechanics,
  and Astronomy}\ }\textbf {\bibinfo {volume} {61}},\ \bibinfo {eid} {31011}
  (\bibinfo {year} {2018})},\ \Eprint {http://arxiv.org/abs/1710.06065}
  {arXiv:1710.06065 [astro-ph.HE]} \BibitemShut {NoStop}%
\bibitem [{\citenamefont {{Yuan}}\ \emph {et~al.}(2022)\citenamefont {{Yuan}},
  \citenamefont {{Zhang}}, \citenamefont {{Chen}},\ and\ \citenamefont
  {{Ling}}}]{2022hxga.book...86Y}%
  \BibitemOpen
  \bibfield  {author} {\bibinfo {author} {\bibfnamefont {W.}~\bibnamefont
  {{Yuan}}}, \bibinfo {author} {\bibfnamefont {C.}~\bibnamefont {{Zhang}}},
  \bibinfo {author} {\bibfnamefont {Y.}~\bibnamefont {{Chen}}}, \ and\ \bibinfo
  {author} {\bibfnamefont {Z.}~\bibnamefont {{Ling}}},\ }in\ \href {\doibase
  10.1007/978-981-16-4544-0_151-1} {\emph {\bibinfo {booktitle} {Handbook of
  X-ray and Gamma-ray Astrophysics. Edited by Cosimo Bambi and Andrea
  Santangelo}}}\ (\bibinfo {year} {2022})\ p.~\bibinfo {pages} {86}\BibitemShut
  {NoStop}%
\end{thebibliography}%

\end{document}